\documentclass[acmsmall, manuscript, nonacm]{acmart}

\AtBeginDocument{%
  }

\usepackage{tabularx}
\usepackage{multirow}
\usepackage{siunitx}
\usepackage{threeparttable}

\usepackage{algorithm}
\usepackage{algpseudocode}
\usepackage{bm}

\newcommand{\PhysDerivatives}[1]{\text{PhysDerivatives}(#1)}
\newcommand{\textgeom}{\text{geom}}

\newcommand{\nDofs}{\text{nDofs}}
\newcommand{\numPoints}{\text{numPoints}}
\newcommand{\InterpolateGradAtPoint}[3]{\mathrm{InterpolateGradAtPoint}(#1, #2, #3)}
\newcommand{\CalculateStress}[3]{\mathrm{CalculateStress}(#1, #2, #3)}
\newcommand{\TransformAndWeight}[3]{\mathrm{TransformAndWeight}(#1, #2, #3)}
\newcommand{\GetBasisGradient}[2]{\mathrm{GetBasisGradient}(#1, #2)}
\newcommand{\Contract}[2]{\mathrm{Contract}(#1, #2)}

\begin{document}

\title{Shifting the Sweet Spot: High-Performance Matrix-Free Method for High-Order Elasticity}

\author{Dali Chang}
\email{changdali@mail.dlut.edu.cn}
\orcid{0009-0009-2565-1823}
\affiliation{%
  \institution{Dalian University of Technology}
  \city{Dalian}
  \state{Liaoning}
  \country{China}
}
\affiliation{%
  \institution{Greater Bay Area National Center of Technology Innovation}
  \city{Guangzhou}
  \state{Guangdong}
  \country{China}
}

\author{Chong Zhang}
\email{zhangchong2020@iscas.ac.cn}
\affiliation{%
  \institution{Institute of Software, Chinese Academy of Sciences}
  \city{Beijing}
  \country{China}
}
\affiliation{%
  \institution{University of Chinese Academy of Sciences}
  \city{Beijing}
  \country{China}
}

\author{Kaiqi Zhang}
\email{zhangkq@mail.dlut.edu.cn}
\affiliation{%
  \institution{Dalian University of Technology}
  \city{Dalian}
  \state{Liaoning}
  \country{China}
}
\affiliation{%
  \institution{Greater Bay Area National Center of Technology Innovation}
  \city{Guangzhou}
  \state{Guangdong}
  \country{China}
}

\author{Mingguan Yang}
\email{yangmingguan@ncti-gba.cn}
\affiliation{%
  \institution{Greater Bay Area National Center of Technology Innovation}
  \city{Guangzhou}
  \state{Guangdong}
  \country{China}
}

\author{Huiyuan Li}
\authornote{Corresponding authors.}
\email{huiyuan@iscas.ac.cn}
\orcid{0000-0002-6326-9926}
\affiliation{%
  \institution{Institute of Software, Chinese Academy of Sciences}
  \city{Beijing}
  \country{China}
}

\author{Weiqiang Kong}
\authornotemark[1]
\email{wqkong@dlut.edu.cn}
\affiliation{%
  \institution{Dalian University of Technology}
  \city{Dalian}
  \state{Liaoning}
  \country{China}
}

\renewcommand{\shortauthors}{Chang et al.}

\begin{abstract}
MFEM is a widely used finite-element library, but its native linear-elasticity Partial Assembly (PA) path still applies an $O((p+1)^6)$ contraction in the element operator, leaving the CPU operator-throughput sweet spot near $p\approx 2$ in our baseline measurements. This work closes this implementation gap for MFEM linear elasticity on affine tensor-product hexahedral meshes by integrating four well-established tensor-product PA optimizations (sum factorization, Voigt notation, macro-kernel fusion, and slice-wise loop reorganization) into MFEM's native linear-elasticity PA path. The resulting operator is evaluated in high-order GMG-PCG solves using MFEM's geometric multigrid (GMG) components. On AMD~EPYC~7713, the optimized operator achieves $7\text{--}83\times$ kernel speedup and $3.6\text{--}16.8\times$ end-to-end speedup across $p\in\{1,2,4,8\}$. At fixed problem size, the kernel-time operator throughput peaks around $p=6$ and remains high at $p=8$, shifting the operator-throughput sweet spot to $p\ge 6$. The same trend is reproduced on Huawei~Kunpeng~920 (ARMv8.2).
These results are accompanied by per-stage ablation and hardware-counter characterization; the implementation will be released on GitHub.
\end{abstract}



\maketitle

\section{Introduction}
\label{sec:introduction}

High-order finite element methods (FEM) achieve high accuracy with relatively few degrees of freedom under regularity assumptions, which makes them attractive for the discretization of partial differential equations in applications such as linear elasticity \cite{karniadakis2013spectral, deville2002high, orszag1979spectral}. The finite element discretization yields a large sparse linear system $\mathbf{A}\mathbf{x} = \mathbf{b}$, and iterative solvers repeatedly apply the matrix-vector product, or more generally the discrete operator, making this operation a central target for reducing per-iteration cost.

Several strategies can implement the matrix-vector product, ranging from Full Assembly (FA) and Element Assembly (EA) to Partial Assembly (PA) and fully unassembled (UA) operator evaluation; Section~\ref{sec:background} defines them through the MFEM operator chain. We use ``matrix-free'' in the broad sense of applying the operator without forming assembled DoF-to-DoF matrices, whether global sparse or element-local dense. PA and UA are matrix-free in this sense, whereas EA is an element-level assembled representation; PA is the concrete representation optimized in this work because the operators are reused across many iterative-solver applications, so setup cost can be amortized, and because PA is a mature production mode in MFEM and libCEED \cite{anderson2021mfem,brown2021libceed}. Among these strategies, the key contrast for this work is between FA and PA. FA stores the global stiffness matrix; for tensor-product elements, the dense element-local matrix has $O((p+1)^{2d})$ entries, and after assembly each scalar row couples to $O((p+1)^d)$ neighboring basis functions, so assembled storage grows polynomially in $p$ and limits the problem sizes that fit on a single node. PA computes the action of the matrix on a vector ``on the fly'', trading per-iteration arithmetic for a reduced memory footprint and exposing the operator to on-chip cache and SIMD optimization \cite{ljungkvist2014matrix}. For tensor-product elements, sum factorization reduces the per-element operator-application cost from $O((p+1)^{2d})$ to $O((p+1)^{d+1})$, i.e., from $O((p+1)^6)$ to $O((p+1)^4)$ for hexahedra in 3D \cite{cao2025towards, wang2024novel, ljungkvist2017matrix, kronbichler2012generic}.

In practice, translating this tensor-product advantage to vector-valued elasticity on multi-core CPUs remains implementation-dependent. Much high-performance PA work has focused on scalar problems such as Poisson operators \cite{ljungkvist2017finite, kronbichler2019fast, kronbichler2019multigrid}; elasticity adds vector components, symmetric stress/strain structure, and contraction of the elasticity tensor $C_{ijkl}$, motivating recent matrix-free elasticity work in deal.II and ExaDG \cite{davydov2020matrix, schussnig2025matrix}. GPU-oriented solid-mechanics studies have also reported substantial gains on accelerators \cite{georgescu2013gpu,brown2022performance}, while CPU-focused elasticity remains sensitive to the operator dataflow and target architecture. Within MFEM, the native v4.8 linear-elasticity PA path studied here provides a general-purpose matrix-free implementation but does not exploit the tensor-product sum-factorization and element-local dataflow optimizations evaluated in this work. Prior libCEED-based CPU elasticity work reports problem-dependent sweet spots, often at lower orders \cite{mehraban2021matrix}. We therefore ask whether the kernel-time operator-throughput sweet spot of MFEM's linear-elasticity PA path can be shifted to high orders ($p \ge 6$). Unless otherwise stated, ``sweet spot'' refers to fixed-problem-size matrix-free \texttt{AddMult} throughput, not an accuracy-per-cost optimum or the polynomial degree minimizing end-to-end GMG-PCG time.

We evaluate the regime targeted by the present implementation: smooth linear elasticity on structured or block-structured affine hexahedral meshes, where tensor-product PA and per-element constant geometry factors can be exploited efficiently. Problems dominated by geometric singularities, material discontinuities, thin features, or all-hex meshing constraints may still favor low-order adaptive discretizations \cite{owen1998survey,piegl2012nurbs}.

An efficient operator must also be evaluated in its solver context \cite{balay2019petsc}. Because PA does not store a global sparse fine-level matrix by default, algebraic preconditioners such as AMG and incomplete factorizations cannot be applied directly to the operator \cite{falgout2002hypre}. This work uses GMG with matrix-free Chebyshev--Jacobi smoothing on fine and intermediate levels and an assembled AMG-preconditioned coarse solve. This combines naturally with the PA method and is an established practice \cite{brown2022performance, kronbichler2019multigrid, clevenger2020flexible}. The core objective is to integrate sum factorization, Voigt notation, kernel fusion, and slice-wise loop reorganization into the MFEM linear-elasticity PA path and to evaluate the resulting operator inside this GMG solver.

\paragraph{Related work.}
Sum factorization for tensor-product elements has roots in spectral and spectral-element methods \cite{orszag1979spectral,deville2002high} and underpins high-order matrix-free implementations on CPUs and GPUs \cite{kronbichler2012generic,kronbichler2019fast,abdelfattah2021gpu}. Recent deal.II/ExaDG solid-mechanics work has demonstrated matrix-free elasticity operators and GMG for finite-strain formulations \cite{davydov2020matrix,schussnig2025matrix,shakeri2024stable}. The matrix-free + GMG composition used here follows the standard pattern of Kronbichler and Ljungkvist \cite{kronbichler2019multigrid}, the deal.II step-37 tutorial \cite{dealii_step37}, and Clevenger et al.~\cite{clevenger2020flexible}. Our focus is to optimize MFEM's native linear-elasticity PA path using established tensor-product and dataflow techniques, characterize the resulting operator, and evaluate it in a GMG-PCG solver context \cite{anderson2021mfem,arndt2023deal,brown2021libceed}.

To close this gap, this work contributes the following for MFEM linear-elasticity PA on tensor-product hexahedra:
\begin{enumerate}
    \item \textbf{An optimized PA operator for MFEM's native linear-elasticity path.} This work integrates sum factorization~\cite{deville2002high,kronbichler2012generic}, Voigt notation~\cite{bhavikatti2005finite}, macro-kernel fusion, and a slice-wise loop organization inspired by tensor-product cache-local evaluation patterns~\cite{kronbichler2019fast} into the native MFEM linear-elasticity PA path. Evaluated within a common MFEM GMG-PCG solver setup for high-order solves, the optimized operator achieves $7\text{--}83\times$ kernel speedup and $3.6\text{--}16.8\times$ end-to-end speedup over the MFEM~v4.8 baseline across $p\in\{1,2,4,8\}$ on AMD~EPYC~7713, shifting the operator-throughput sweet spot to $p\ge 6$.

    \item \textbf{Performance characterization of the optimized operator.} A per-stage timing ablation traces how the dominant bottleneck shifts from arithmetic work through memory bandwidth to microarchitectural effects as the four optimizations are applied in sequence. A hardware-counter characterization on AMD~EPYC~7713 combines theoretical FLOPs/DoF accounting with LIKWID-measured operational intensity and sustained DRAM bandwidth, and Roofline placement at two problem scales (6.5M and 51.17M DoFs) shows that the bandwidth-bound regime is preserved across scales.

    \item \textbf{Portability and context across CPU matrix-free elasticity work.} Replaying the same optimization stack on Huawei~Kunpeng~920 (ARMv8.2) reproduces the operator-throughput sweet-spot shift observed on AMD~EPYC~7713. A cross-paper comparison reports FLOPs/DoF and sustained DRAM utilization alongside raw throughput, placing PAop in context with deal.II-based elasticity codes~\cite{schussnig2025matrix}.
\end{enumerate}

The remainder of this work is organized as follows: background and PA discretization (Section~\ref{sec:background}), the GMG-PCG solver configuration (Section~\ref{sec:gmg_architecture}), the four-stage operator optimization (Section~\ref{sec:optimization}), experiments and cross-paper comparison (Section~\ref{sec:experiments}), and conclusion (Section~\ref{sec:conclusion}).

\section{Background: Mathematical Model and Numerical Discretization}
\label{sec:background}

This section summarizes the linear-elasticity discretization, MFEM's operator decomposition $A = P^T G^T B^T D B G P$, and the tensor-product sum factorization used in Section~\ref{sec:optimization}. The operator-chain view also fixes the FA / EA / PA / UA terminology used throughout the experiments \cite{anderson2021mfem,andrej2024high,brown2021libceed}.

\subsection{Governing Equations and Finite Element Discretization}
\label{ssec:governing_equations}

The physical behavior of a deformable body under external forces, assuming small displacements, is governed by the equations of linear elasticity. The strong form of the static equilibrium problem, defined on a domain $\Omega$ with boundary $\Gamma = \Gamma_D \cup \Gamma_N$, is
\begin{equation}
-\nabla \cdot \boldsymbol{\sigma}(\mathbf{u}) = \mathbf{f} \quad \text{in } \Omega,\qquad
\mathbf{u} = \mathbf{0} \text{ on } \Gamma_D,\qquad
\boldsymbol{\sigma}(\mathbf{u})\cdot\mathbf{n} = \mathbf{t} \text{ on } \Gamma_N,
\label{eq:elasticity_strong} \tag{1}
\end{equation}
where $\mathbf{u}$ is the displacement field, $\mathbf{f}$ is the body force vector, $\boldsymbol{\sigma}$ is the Cauchy stress tensor, $\mathbf{n}$ is the outward unit normal on $\Gamma_N$, and $\mathbf{t}$ is the prescribed Neumann traction. The benchmark material parameters and load values are given in Section~\ref{sssec:benchmark}. The stress is related to the strain $\boldsymbol{\varepsilon}(\mathbf{u}) = \frac{1}{2}(\nabla \mathbf{u} + (\nabla \mathbf{u})^T)$ through a constitutive model. For a general anisotropic material, this is a linear relationship defined by a fourth-order elasticity tensor $\mathbf{C}$. In component form, this relationship is expressed as:
\begin{equation}
\sigma_{ij} = \sum_{k=1}^{3} \sum_{l=1}^{3} \mathbf{C}_{ijkl} \varepsilon_{kl} \label{eq:anisotropic} \tag{2}
\end{equation}
However, for the widely applicable case of isotropic materials, this relationship simplifies to Hooke's Law:
\begin{equation}
\sigma(\mathbf{u}) = \lambda (\nabla \cdot \mathbf{u}) \mathbf{I} + 2\mu \boldsymbol{\varepsilon}(\mathbf{u}) \label{eq:hookes_law} \tag{3}
\end{equation}
where $\lambda$ and $\mu$ are the Lamé material parameters and $\mathbf{I}$ is the identity tensor \cite{bhavikatti2005finite}.

The weak form is: find $\mathbf{u} \in V$ such that for all test functions $\mathbf{v} \in V$,
\begin{equation}
\underbrace{\int_{\Omega} \sigma(\mathbf{u}) : \boldsymbol{\varepsilon}(\mathbf{v}) \,d\Omega}_{a(\mathbf{u},\mathbf{v})} = \underbrace{\int_{\Omega} \mathbf{f} \cdot \mathbf{v} \,d\Omega + \int_{\Gamma_N} \mathbf{t} \cdot \mathbf{v} \,d\Gamma}_{L(\mathbf{v})} \label{eq:weak_form} \tag{4}
\end{equation}
Here, $V$ is a suitable Sobolev space, $a(\mathbf{u}, \mathbf{v})$ is the internal virtual work, and $L(\mathbf{v})$ is the external virtual work. Since $\sigma(\mathbf{u})$ is symmetric, $\sigma(\mathbf{u}):\boldsymbol{\varepsilon}(\mathbf{v})=\sigma(\mathbf{u}):\nabla\mathbf{v}$, the form used by the operator kernels below.

Discretizing $\Omega$ with $H^1$-conforming finite elements gives $\mathbf{u}_h = \sum_{j=1}^{N} \mathbf{U}_j \phi_j$, where $\{\phi_j\}$ are basis functions and $\mathbf{U}$ is the vector of unknown nodal displacements. Galerkin discretization of Equation~\eqref{eq:weak_form} yields the sparse linear system
\begin{equation}
\mathbf{A}\mathbf{U}=\mathbf{F} \label{eq:linear_system} \tag{5}
\end{equation}
where $\mathbf{A}$ is the global stiffness matrix and $\mathbf{F}$ is the global load vector. The entries of $\mathbf{A}$ are assembled from elemental contributions $a(\phi_i,\phi_j)$ \cite{karniadakis2013spectral, deville2002high, orszag1979spectral}. We use the standard six-component Voigt representation for symmetric stress and strain tensors; its computational role in the optimized kernel is described in Section~\ref{ssec:voigt_optimization}.

For the constrained, positive-parameter linear-elasticity problems considered here, the resulting global stiffness matrix $\mathbf{A}$ is sparse, symmetric, and positive-definite (SPD). For large-scale problems, such systems are typically solved by iterative methods such as the preconditioned Conjugate Gradient (PCG) algorithm. In a matrix-based PCG, the sparse matrix--vector product (SpMV) is the recurring operator application; in a matrix-free PCG, this role is filled by on-the-fly evaluation of the operator chain. Accordingly, Section~\ref{ssec:hpc_strategies} focuses on operator-application strategies, while solver-level timings and iteration counts for the preconditioned runs are reported in Section~\ref{ssec:preconditioner_perf} \cite{kronbichler2023fast}.

\subsection{High-Performance Computing Strategies for FEM Operators}
\label{ssec:hpc_strategies}

\begin{figure}[htbp]
    \centering
    \includegraphics[width=\linewidth]{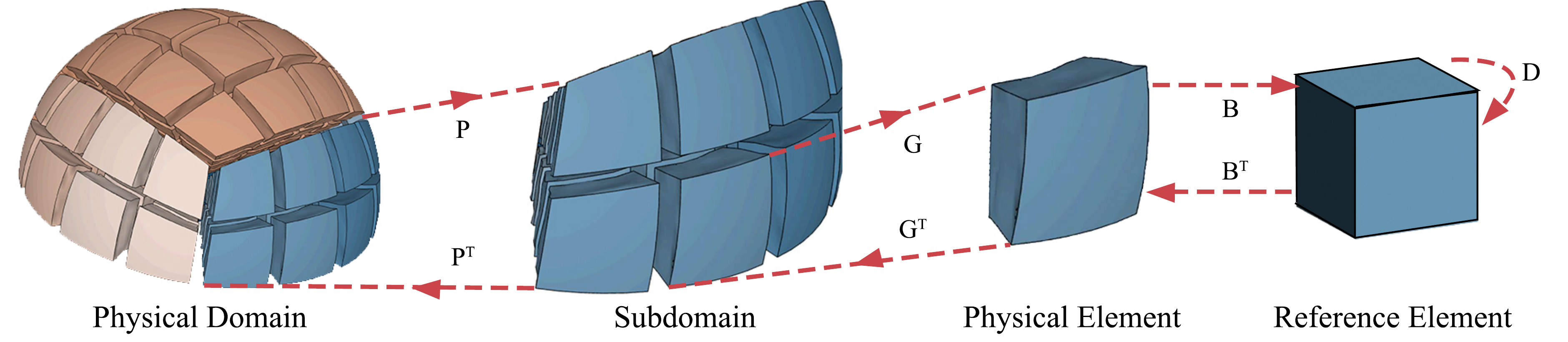}
    \caption{Conceptual decomposition of the FEM operator into a chain of discrete operators, as implemented in libraries like MFEM.}
    \Description{Diagram of the MFEM operator chain from global degrees of freedom through element restriction, basis operations, and quadrature-point data.}
    \label{fig:operator_chain}
\end{figure}

MFEM represents the global discrete operator as a chain $A = P^T G^T B^T D B G P$ (Figure~\ref{fig:operator_chain}) \cite{anderson2021mfem,andrej2024high}. In this chain, \(P\) maps global degrees of freedom to subdomain-local degrees of freedom, \(G\) maps them to element-local degrees of freedom, \(B\) applies basis interpolation and gradients at quadrature points, and \(D\) stores the quadrature-point physics, geometry, and weights. The choice of how far to collapse this chain into stored matrices, whether FA, EA, PA, or fully unassembled (UA) evaluation, determines the per-iteration cost and memory footprint: FA stores the global sparse DoF-to-DoF matrix, EA stores dense element-local DoF-to-DoF matrices, PA stores the quadrature-point operator data \(D\), and UA recomputes \(D\) during each operator application.

\subsubsection{Full Assembly}
\label{sssec:fa}

Full Assembly precomputes and stores the global stiffness matrix~$\mathbf{A}$ before the solution stage by iterating over every mesh element, computing each element-local matrix, and assembling the contributions into a single global sparse matrix. The subsequent iterative solve then reduces each matrix--vector multiplication to a standard sparse matrix--vector product (SpMV). For high-order finite elements, however, this strategy faces two well-known limitations.

First, capacity. The dense element-local matrix has $O((p+1)^{2d})$ entries, and after assembly each scalar row of $\mathbf{A}$ couples to $O((p+1)^d)$ neighboring basis functions; total assembled storage therefore grows polynomially in $p$. On a distributed-memory parallel machine the matrix is partitioned across nodes, but the per-node memory still constrains how large a problem fits at a given polynomial degree, and at high $p$ this constraint becomes the dominant limit; the empirical out-of-memory thresholds in our experiments are reported in Section~\ref{ssec:macro_perf}.

Second, bandwidth. SpMV is a well-studied memory-bound kernel: its arithmetic intensity is low because each nonzero entry of $\mathbf{A}$ is read from main memory once but used in only a single multiply--add \cite{chaudhuri2008computer,gropp1999toward}. Cache reuse is further limited by the indirect indexing through column-index arrays, which in unstructured meshes implies a non-stride-one access into the input vector $\mathbf{u}$. In Roofline terms, this low arithmetic intensity places SpMV below the machine balance point, so performance is limited by the bandwidth roof even when memory bandwidth is used efficiently \cite{williams2009roofline}. This observation motivates operator representations that reduce global-memory traffic by spending additional local arithmetic.

\subsubsection{Partial Assembly as the Matrix-Free Representation Used in This Work}
\label{sssec:pa}

In the terminology used here, matrix-free methods apply the operator without forming assembled DoF-to-DoF matrices, whether global sparse or element-local dense \cite{carey1986element, carey1988element, ljungkvist2017finite}. Partial Assembly (PA) is the concrete matrix-free representation studied in this work: it precomputes and stores the quadrature-point operator data $D$, which combines quadrature weights, geometric (Jacobian-derived) factors, and the material/constitutive data, while the restriction and basis actions around $D$ are applied on the fly. Fully unassembled (UA) variants would also recompute $D$ during each operator application, reducing setup storage further at the cost of repeated geometry and material evaluation. Both PA and UA avoid assembled DoF-to-DoF matrices; this work chooses PA because the same operator is applied many times inside PCG and GMG, allowing the stored quadrature data to be amortized. PA therefore follows the compute-for-memory tradeoff identified above: it avoids repeatedly streaming the assembled sparse matrix and instead evaluates the remaining restriction and basis actions on the fly. A naive 3D tensor-product PA application would make this added arithmetic scale as $O(p^6)$ per element, so sum factorization is needed to keep the tradeoff favorable \cite{kronbichler2012generic}.

\subsubsection{Tensor-Product Sum Factorization}
\label{sssec:sum_factorization}

This work focuses on tensor-product quadrilateral and hexahedral elements; efficient matrix-free methods for simplicial elements follow a separate line of work \cite{moxey2020}. A 3D tensor-product basis function $\Phi(\xi, \eta, \zeta)$ can be expressed as the product of three 1D basis functions $\phi(r)$:
\begin{equation}
\Phi_{ijk}(\xi, \eta, \zeta) = \phi_i(\xi) \cdot \phi_j(\eta) \cdot \phi_k(\zeta) \label{eq:tensor_basis} \tag{6}
\end{equation}
This structure allows us to decompose high-dimensional operators into a series of operations on 1D components. Sum factorization is the technique that uses this structure to reformulate a single, high-dimensional tensor contraction into an equivalent sequence of lower-dimensional contractions on higher-rank tensors. Conceptually, the action of a 3D gradient operator, ${G}_{3D}$, can be decomposed into the sum of contributions from the three spatial directions. For example, the partial derivative operator in the $\xi$ direction, ${G}_{\xi}$, can be represented (up to the chosen vectorization convention) as a Kronecker product of 1D operator matrices: $G_{\xi}=\mathbf{G}_{1D}\otimes\mathbf{B}_{1D}\otimes\mathbf{B}_{1D}$, where $\mathbf{G}_{1D}$ is the 1D derivative operator and $\mathbf{B}_{1D}$ is the 1D interpolation operator (not to be confused with the element-restriction operator $G$ of the operator chain). The essence of sum factorization is that instead of explicitly forming and storing the large dense matrices defined by the Kronecker products, their action is implemented equivalently through a sequence of 1D matrix multiplications \cite{deville2002high, orszag1979spectral, abdelfattah2021gpu}. It is this ``factorized evaluation'' strategy that reduces the complexity of applying a local operator from $O(p^{2d})$ to $O(d \cdot p^{d+1})$, which in 3D corresponds to $O(p^6) \to O(p^4)$. The cache and SIMD implications of this reorganization are what Section~\ref{sec:optimization} exploits for the linear-elasticity PA kernel.

\section{Configuration of the Geometric Multigrid Preconditioner}
\label{sec:gmg_architecture}

This section describes the GMG preconditioner used with the PA and PAop operators. The architecture follows the matrix-free + algebraic-coarse-solve template established by Kronbichler and Ljungkvist \cite{kronbichler2019multigrid} and Clevenger et al.\ \cite{clevenger2020flexible}, and is constructed from MFEM's built-in multigrid, finite-element-space hierarchy, Chebyshev smoother, and HypreBoomerAMG components \cite{anderson2021mfem}. Section~\ref{ssec:preconditioner_perf} reports the per-phase cost breakdown for this configuration.

\noindent\textbf{Hierarchy.}~The multigrid levels are built on a single parallel mesh. Starting from the coarse mesh, $r$ uniform geometric refinements yield levels $\ell = 0, 1, \dots, r$ at polynomial degree $p_{\min}=1$, after which $p$-refinements double the polynomial degree until the finest level reaches $p \in \{1,2,4,8\}$. Each level owns its own $H^1$ continuous Galerkin space.

\noindent\textbf{Transfer operators.}~Inter-grid transfer uses MFEM's standard \texttt{ParFiniteElementSpaceHierarchy} (natural injection for $h$-refined levels, polynomial interpolation for $p$-refined levels).

\noindent\textbf{Boundary conditions.}~Dirichlet conditions are applied per level using the same boundary attribute as the finest level.

A standard $V$-cycle consists of three components: a smoother on the fine and intermediate levels, the inter-grid transfer operators, and a coarse-grid solver. The GMG configuration uses a matrix-free smoother on the computationally intensive fine and intermediate levels, while switching to a matrix-based solution strategy on the small-scale coarsest level (Sections~\ref{ssec:chebyshev_smoother} and \ref{ssec:coarse_solver}).

\subsection{Smoother on Fine and Intermediate Levels}
\label{ssec:chebyshev_smoother}

A key component of the multigrid cycle is the smoother, whose responsibility is to damp high-frequency error. Classical relaxations such as Gauss--Seidel or SOR require row-wise matrix access, which conflicts with the matrix-free representation used on fine and intermediate levels. We therefore use MFEM's \texttt{OperatorChebyshevSmoother}, a Chebyshev-accelerated Jacobi smoother used in matrix-free multigrid practice \cite{kronbichler2019multigrid,clevenger2020flexible,dealii_step37,phillips2025chebyshev}. It requires only the action of $\bm{A}$ and the diagonal of $\bm{A}$; in our PA setting, the action is the kernel of Section~\ref{sec:optimization}, and the diagonal is obtained from \texttt{BilinearForm::AssembleDiagonal}.

The smoother applies a polynomial $p_k(\bm{D}^{-1}\bm{A})$ to the residual, where $\bm{D}$ is the operator diagonal. We use Chebyshev degree $k=2$, estimate $\lambda_{\max}$ with $10$ power iterations, and apply one pre-smoothing and one post-smoothing step per $V(1,1)$ cycle. All fine-level variants use the same smoother semantics; only the operator handle and, for FA, the diagonal source differ.

\subsection{Coarse-Grid Solver and Stopping Criteria}
\label{ssec:coarse_solver}

The coarsest level contains the fewest degrees of freedom in the $V$-cycle, so assembling only its sparse stiffness matrix keeps the matrix-based setup confined to the smallest system while enabling the AMG-preconditioned inexact coarse solve that supplies the GMG coarse-grid correction. We solve this coarsest-level system with PCG preconditioned by \texttt{HypreBoomerAMG} \cite{falgout2002hypre}. BoomerAMG is configured for vector-valued elasticity using the systems option with vector dimension $3$; the effective strong threshold is $0.5$, and the remaining relaxation, coarsening, and interpolation options use MFEM defaults. Fine and intermediate levels remain matrix-free, and Section~\ref{ssec:preconditioner_perf} quantifies the resulting GMG configuration's effect on iteration counts and solve time.

The coarse solve is inexact: the inner PCG uses $(\text{rel\_tol}=\sqrt{10^{-4}},\,\text{abs\_tol}=0,\,\text{max\_iter}=10)$ with \texttt{HypreBoomerAMG} as preconditioner. This inner solve is embedded in the outer PCG described below.

\begin{figure}[htbp]
    \centering
    \includegraphics[width=0.6\linewidth]{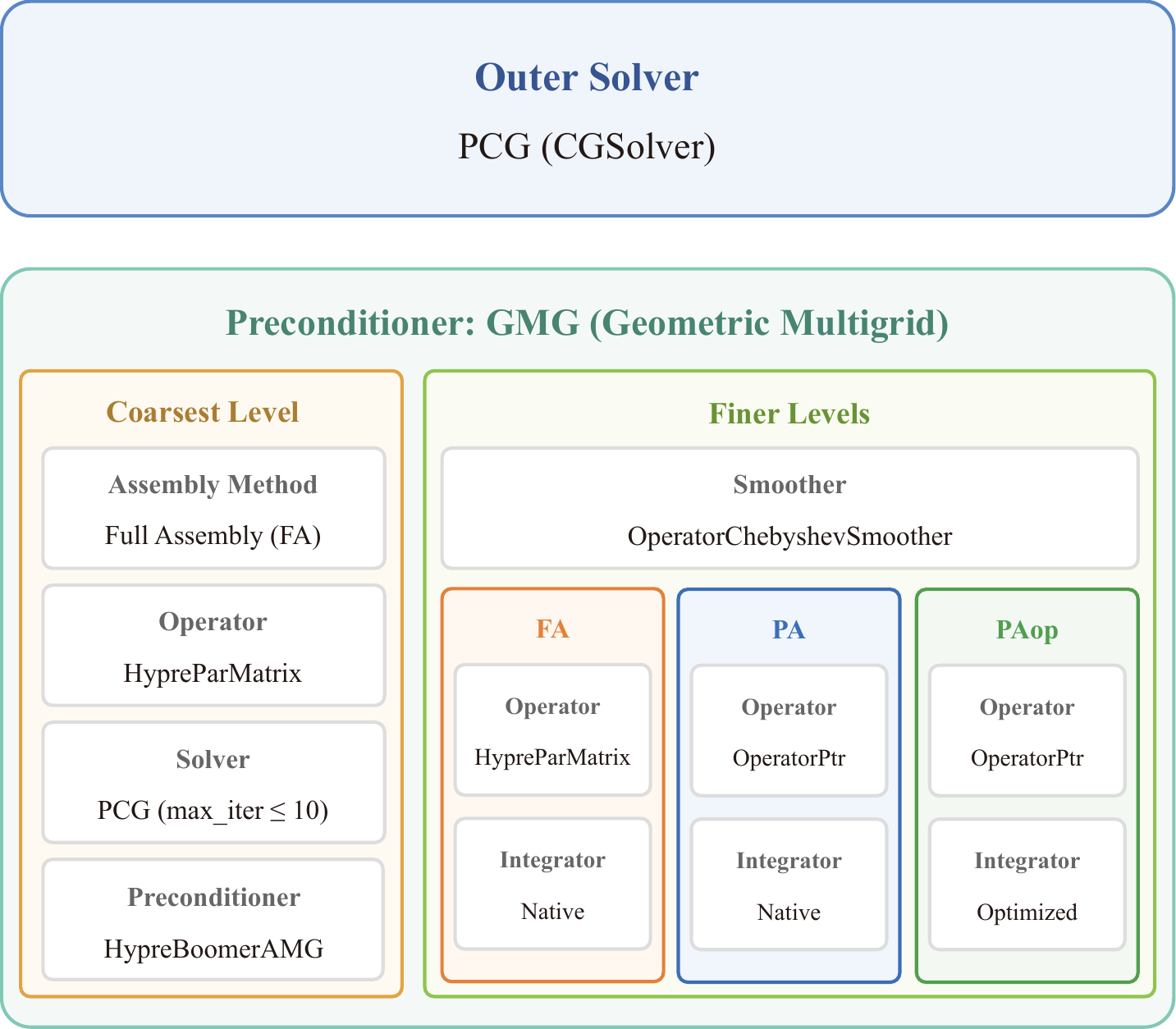}
    \caption{Hybrid GMG preconditioner. Fine and intermediate levels keep the operator in matrix-free PA/PAop form with Chebyshev$(k{=}2)$--Jacobi smoothing; the coarsest level is assembled and solved by inexact PCG-AMG. FA+GMG, PA+GMG, and PAop+GMG differ only in the operator type used on fine and intermediate levels.}
    \Description{Schematic of the geometric multigrid hierarchy with matrix-free fine levels, Chebyshev-Jacobi smoothing, and an assembled AMG-preconditioned coarse solve.}
    \label{fig:gmg_architecture}
\end{figure}

The complete preconditioner (Figure~\ref{fig:gmg_architecture}) wraps the matrix-free PA fine-level smoother and the AMG-on-the-coarsest-level coarse solve described above. In all experiments, the outer PCG uses relative tolerance $10^{-6}$; for preconditioned solves MFEM tests $(B r_k,r_k)^{1/2}/(B r_0,r_0)^{1/2}\le10^{-6}$, where $B$ denotes the GMG preconditioner, with a $5{,}000$-iteration cap (never reached in the runs we report).

\section{Four-Stage Optimization of the Linear-Elasticity PA Operator}
\label{sec:optimization}

This section details the four-stage optimization stack we apply to MFEM's baseline linear-elasticity Partial Assembly operator (\texttt{ElasticityIntegrator}, MFEM~v4.8). The four stages are: (i) macro-kernel fusion that merges gradient interpolation, stress evaluation, and the operator action into one element-local kernel (Section~\ref{ssec:macro_kernel_fusion}); (ii) Voigt notation that replaces the symmetric $3\times 3$ stress and strain by their six independent components (Section~\ref{ssec:voigt_optimization}); (iii) sum factorization that decomposes the per-element contraction into a sequence of one-dimensional contractions and applies the geometry transformation in-line (Section~\ref{ssec:sum_factorization_algo}); and (iv) a slice-wise reorganization of the contraction loops that keeps the working set in L1/L2 cache (Section~\ref{ssec:loop_optimization}).

Sections~\ref{ssec:macro_kernel_fusion}--\ref{ssec:loop_optimization} are organized around the cost removed at each stage. Macro-kernel fusion removes the operator-wide \texttt{QVec} round trip, Voigt notation removes redundant symmetric-tensor storage and arithmetic, sum factorization replaces dense element contractions by one-dimensional tensor contractions, and the slice-wise loop organization bounds the element-local working set. The last two stages are described along the forward/backward dataflow of the final kernel: Section~\ref{ssec:sum_factorization_algo} presents the forward tensor contractions and the intermediate \texttt{sm1} required by the subsequent $Z$ contraction, while Section~\ref{ssec:loop_optimization} analyzes the slice-wise organization, including the cross-slice role of \texttt{sm1}, together with pointwise stress evaluation and the backward transpose contractions.

In practice, the choice of~$p$ is a trade-off between application demands and computational cost: engineering workflows often use lower orders ($p=1$ or $p=2$ \cite{gokhale2008practical}) to control cost, whereas high-order research explores larger orders (e.g., $7 \leq p \leq 11$ \cite{karp2022high}) in pursuit of efficiency. We select $p\in\{1,2,4,8\}$ as the core range because it covers both the low-order baseline operator-throughput sweet spot and the high-order regime where the optimized PA operator is expected to become advantageous.

\subsection{Baseline Performance and Bottleneck Analysis}
\label{ssec:baseline_analysis}

A profiler-guided analysis of the MFEM~v4.8 linear-elasticity PA operator with Intel VTune Profiler and Linux perf \cite{de2010new} identifies the matrix-vector multiplication function \texttt{ElasticityAddMultPA} as the dominant hotspot, accounting for over $90\%$ of execution time. The logic of this baseline operator (Algorithm~\ref{alg:baseline_operator_compact}) splits its core task $\mathbf{y} += A(\mathbf{x})$ into two computational kernels. The first kernel iterates over all quadrature points to compute the stress tensor and writes the results to an intermediate array \texttt{QVec}. The second kernel then iterates over all elements again, reading back the entire \texttt{QVec} array to perform a tensor contraction and accumulate the result. This two-stage layout has two performance costs. First, it imposes a memory round-trip: the results of Kernel~1 are written to main memory (DRAM) and read back in full by Kernel~2, destroying temporal locality and dominating the operator's memory traffic. Second, Kernel~2 incurs a high arithmetic cost from a high-dimensional tensor contraction, $Y_{iqe} += \sum_{\alpha,m} \text{QVec}_{\alpha mqe} \cdot G_{\alpha mi}$, which has computational complexity $O(p^{2d})$ and requires streaming the full basis gradient matrix $G$ from main memory.

\begin{algorithm}[h]
    \caption{Baseline MFEM PA \texttt{AddMult} dataflow with an operator-wide \texttt{QVec} round trip}
    \label{alg:baseline_operator_compact}
    \begin{algorithmic}[1]
        \State \textbf{Input:} Input vector $X$, Lamé parameters $\lambda, \mu$, geometry info $\text{geom}$ \quad \textbf{Output:} Output vector $Y$
        
        \State $\text{QVec}_{\nabla u} \leftarrow \PhysDerivatives(X)$ \Comment{Compute physical gradients}
        
        \State // Kernel 1: Compute geometrically-transformed stress at quadrature points
        \For{each element $e$}
            \For{each quadrature point $\alpha$} \Comment{Load data for the current point}
                \State $\nabla u \leftarrow \text{QVec}_{\nabla u}(\alpha, e)$; \quad $J \leftarrow \text{geom.J}(\alpha, e)$; \quad $w_\alpha \leftarrow \text{ipWeights}[\alpha]$
                \State $\bm{\varepsilon}(u) \leftarrow \frac{1}{2}(\nabla u + (\nabla u)^T)$; \quad $\bm{\sigma}(u) \leftarrow \lambda(\alpha,e) (\nabla \cdot u) I + 2\mu(\alpha,e) \bm{\varepsilon}(u)$
                \State $\text{QVec}_{\sigma}(\alpha, e) \leftarrow w_\alpha \det(J) \cdot \big(J^{-1} \bm{\sigma}(u)\big)$
            \EndFor
        \EndFor

        \State // Kernel 2: Apply operator action (gradient of test function)
        \For{each element $e$}
            \For{$i = 0$ to $\text{nDofs}-1$ and $q = 0$ to $d-1$} \Comment{Loop over test function DOFs}
                \State $s \leftarrow \sum_{\alpha=0}^{\numPoints-1} \sum_{m=0}^{d-1} \text{QVec}_{\sigma}(\alpha, m, q, e) \cdot G(\alpha, m, i)$
                \State $Y(i, q, e) \leftarrow Y(i, q, e) + s$
            \EndFor
        \EndFor
    \end{algorithmic}
\end{algorithm}

To quantify this bottleneck, we use an operational-intensity (OI) model. For a 3D problem ($d=3$) in double precision, the asymptotic OI of the baseline operator is approximately $0.75$~FLOP/Byte; the same computation reaches approximately $0.96$~FLOP/Byte at $p=8$ in our LIKWID measurements on AMD~EPYC~7713 \cite{treibig2010likwid}. The baseline is both high-FLOP and low-OI because redundant interpolation FLOPs coexist with streaming of the full basis-gradient matrix from main memory \cite{kronbichler2023fast}. These two observations motivate the optimizations that follow: macro-kernel fusion removes the operator-wide \texttt{QVec} round trip, whereas sum factorization reduces the dense per-element contraction cost.

\subsection{Macro-kernel Fusion}
\label{ssec:macro_kernel_fusion}

Macro-kernel fusion addresses the memory round-trip bottleneck by consolidating the three baseline stages (Gradient Interpolation, Stress Calculation, and Operator Action) into a single element-centric kernel. The fused dataflow (Algorithm~\ref{alg:fused_kernel_compact}) places the entire operator application within one parallel loop over elements.

\begin{algorithm}[h]
    \caption{Macro-kernel fusion dataflow. This sketch removes the operator-wide \texttt{QVec} array of Algorithm~\ref{alg:baseline_operator_compact}.}
    \label{alg:fused_kernel_compact}
    \begin{algorithmic}[1]
        \State \textbf{Input:} Input vector $X$, Lamé parameters $\lambda, \mu$, geometry info $\text{geom}$ \quad \textbf{Output:} Output vector $Y$
        
        \For{each element $e$}
            \For{$\alpha = 0$ to $\numPoints-1$}
                \State $\nabla u_\alpha \leftarrow \InterpolateGradAtPoint{X}{e}{\alpha}$
                \State $\bm{\sigma}_\alpha \leftarrow \CalculateStress{\nabla u_\alpha}{\lambda_\alpha}{\mu_\alpha}$
                \State $\bm{Q}_\alpha \leftarrow \TransformAndWeight{\bm{\sigma}_\alpha}{\textgeom}{\alpha}$
            \EndFor
            
            \For{$i = 0$ to $\nDofs-1$}
                \State $\bm{Y}_i \leftarrow \sum_{\alpha=0}^{\numPoints-1} \Contract{\bm{Q}_\alpha}{\GetBasisGradient{i}{\alpha}}$
                \State $Y(i, e) \leftarrow \bm{Y}_i$
            \EndFor
        \EndFor
    \end{algorithmic}
    \vspace{0.25em}
    {\footnotesize\itshape\textbf{Note.} The final PAop kernel expands the interpolation and contraction steps into the sum-factorized sweeps of Sections~\ref{ssec:sum_factorization_algo} and~\ref{ssec:loop_optimization}.\par}
\end{algorithm}

For each element, all computational stages are executed sequentially, with intermediate data staged in an on-chip buffer, \texttt{el\_QVec}, sized for only a single element ($8d^2(p+2)^3$~bytes, about $72$~KB for $d=3,\ p=8$). Fusion eliminates global \texttt{QVec} traffic and improves producer-consumer locality; the remaining main-memory traffic is dominated by the input/output vectors and material, geometry, quadrature-weight, and basis data. Merging multiple parallel regions into one also avoids redundant synchronization points. Algorithm~\ref{alg:fused_kernel_compact} sketches this fusion-level dataflow only; the on-chip buffers that complete the PAop kernel, together with their sizes and lifetimes, are summarized in Table~\ref{tab:b1_buffers}.

\subsection{Optimizing Computation and Storage with Voigt Notation}
\label{ssec:voigt_optimization}

After addressing the operator-wide data-movement bottleneck with kernel fusion, we use Voigt notation to improve computational efficiency within the fused kernel. The baseline implementation operates on full $d \times d$ stress and strain tensors, but in linear elasticity, both are symmetric, leading to wasted storage and redundant floating-point operations. Voigt notation provides a lossless mapping from a symmetric $3 \times 3$ tensor to a 6-component vector. We use the implementation ordering $[\sigma_{11},\sigma_{22},\sigma_{33},\sigma_{12},\sigma_{13},\sigma_{23}]^T$ in one-based notation, corresponding to the zero-based buffer order $[00,11,22,01,02,12]$. This six-component constitutive form is implemented with the structured arithmetic in Section~\ref{ssec:loop_optimization} rather than as a dense $6\times6$ matrix-vector product; Equation~\eqref{eq:mult} shows the equivalent isotropic relation.

\begin{equation}
\begin{pmatrix} \sigma_{11} \\ \sigma_{22} \\ \sigma_{33} \\ \sigma_{12} \\ \sigma_{13} \\ \sigma_{23} \end{pmatrix} =
\begin{pmatrix} 
\lambda + 2\mu & \lambda & \lambda & 0 & 0 & 0 \\ 
\lambda & \lambda + 2\mu & \lambda & 0 & 0 & 0 \\ 
\lambda & \lambda & \lambda + 2\mu & 0 & 0 & 0 \\ 
0 & 0 & 0 & \mu & 0 & 0 \\ 
0 & 0 & 0 & 0 & \mu & 0 \\ 
0 & 0 & 0 & 0 & 0 & \mu 
\end{pmatrix} 
\begin{pmatrix} \varepsilon_{11} \\ \varepsilon_{22} \\ \varepsilon_{33} \\ 2\varepsilon_{12} \\ 2\varepsilon_{13} \\ 2\varepsilon_{23} \end{pmatrix}
\label{eq:mult}
\tag{7}
\end{equation}

Voigt notation reduces the per-quadrature-point stress-evaluation work from the baseline $3\times3\times3\times3$ contraction pattern with up to $81$ scalar tensor terms to roughly $24$ floating-point operations in the structured form under our multiply/add counting convention; the structured arithmetic is given in Section~\ref{ssec:loop_optimization}. It also reduces on-chip stress storage from nine to six components and avoids intermediate full $3{\times}3$ matrices in the fused dataflow.

\subsection{Sum Factorization for Forward Gradient Evaluation}
\label{ssec:sum_factorization_algo}

Kernel fusion and Voigt notation reduce dataflow and redundant stress arithmetic, but the per-element contraction cost still grows as $O((p+1)^6)$ when implemented as a dense matrix multiply. We replace the dense contraction by sum factorization \cite{deville2002high,kronbichler2012generic} adapted to the MFEM linear-elasticity PA path. We use $D1D=p+1$ for the 1D DoFs and $Q1D=p+2$ for the 1D Gauss--Legendre quadrature points, with 1D interpolation and derivative tables $B(D1D,Q1D)$ and $G(D1D,Q1D)$ staged in cache. The tables $B$ and $G$ are the quadrature-sampled matrix representations of the one-dimensional operators $\mathbf{B}_{1D}$ and $\mathbf{G}_{1D}$ of Section~\ref{sssec:sum_factorization}. The elements are affinely mapped, so $J^{-1}$ and $\det(J)$ are constant per element and are precomputed once from the geometry data. We first describe the sum-factorized forward sweep, which evaluates the displacement gradient at quadrature points. The transpose contractions use the same one-dimensional $B$ and $G$ tables and are described in Section~\ref{ssec:loop_optimization}. In the forward sweep, the $Y$-contracted intermediate is stored in \texttt{sm1} and carried across $iz$ slices before the $Z$ contraction; Section~\ref{ssec:loop_optimization} analyzes how this same buffer realizes the slice-wise working-set organization.

\paragraph{Forward contractions in $X$, $Y$, $Z$.}
For each vector component $c\in\{0,1,2\}$ and each $z$-DoF index $iz \in \{0,\dots,p\}$, the gradient $\partial_\xi \mathbf{u}, \partial_\eta \mathbf{u}, \partial_\zeta \mathbf{u}$ at the quadrature points is evaluated in three sequential one-dimensional contractions:
\begin{enumerate}
\item \emph{$X$ contraction.} For every fixed $(iy, iz, c)$, compute
\begin{align*}
u(qx) &= \sum_{ix} \mathtt{x}(ix,iy,iz,c,e)\,B(ix,qx),\\
v(qx) &= \sum_{ix} \mathtt{x}(ix,iy,iz,c,e)\,G(ix,qx).
\end{align*}
The two outputs are written to a small two-channel buffer \texttt{sm0[0/1]} indexed by $(iy, qx)$.

\item \emph{$Y$ contraction.} For every fixed $(qx, iz, c)$, contract the previous step's $u, v$ along $iy$:
\begin{align*}
\partial_\xi(qx,qy) &= \sum_{iy} v(iy,qx)\, B(iy,qy),\\
\partial_\eta(qx,qy) &= \sum_{iy} u(iy,qx)\, G(iy,qy),\\
u_{xy}(qx,qy) &= \sum_{iy} u(iy,qx)\, B(iy,qy).
\end{align*}
The three outputs are written to intermediate buffers \texttt{sm1[0/1/2]} indexed by $(iz, qy, qx)$.

\item \emph{$Z$ contraction.} After the $iz$ loop has filled \texttt{sm1} for every $iz$, a single $Z$-pass contracts along $iz$ to evaluate the reference gradient at the quadrature points $(qz, qy, qx)$:
\begin{align*}
\partial_\xi(qx,qy,qz) &= \sum_{iz} \mathtt{sm1}[0](iz,qy,qx)\,B(iz,qz),\\
\partial_\eta(qx,qy,qz) &= \sum_{iz} \mathtt{sm1}[1](iz,qy,qx)\,B(iz,qz),\\
\partial_\zeta(qx,qy,qz) &= \sum_{iz} \mathtt{sm1}[2](iz,qy,qx)\,G(iz,qz).
\end{align*}
The three components are then converted to the physical-space gradient by the constant per-element matrix $J^{-T}$, formed from the precomputed $J^{-1}$:
\[
(\partial_x,\partial_y,\partial_z)\mathbf{u} \;=\; J^{-T} \cdot (\partial_\xi, \partial_\eta, \partial_\zeta)\mathbf{u},
\]
with the result written to the quadrature-point tensor \texttt{grad\_vec}.
\end{enumerate}

\subsection{Slice-Wise Loop Organization and Backward Contraction}
\label{ssec:loop_optimization}

Sum factorization lowers the contraction arithmetic, but the loop organization used to hand off intermediates between one-dimensional contractions still determines the per-element working set. A direct factorized implementation may still materialize full-volume intermediates between one-dimensional tensor-product passes, increasing the on-chip footprint. PAop instead uses a slice-wise loop organization with element-local buffers across the forward and backward sweeps. In this organization, the forward pass carries the $Y$-contracted values in \texttt{sm1}. The buffers \texttt{grad\_vec} and \texttt{stress} remain element-local quadrature-volume buffers, and the backward pass uses short-lived per-output buffers \texttt{tmpZ}/\texttt{tmpY}. The corresponding buffer sizes and lifetimes are summarized in Table~\ref{tab:b1_buffers}. Within this layout, the kernel evaluates the six-component pointwise stress at each quadrature point and applies the backward transpose contractions that accumulate the nodal result.

\paragraph{Cross-slice intermediate and working-set footprint.}
The cross-slice buffer \texttt{sm1[3][Q1D*Q1D*D1D]} is the key data-structure choice in the slice-wise loop organization. It is populated by the forward $Y$ contractions described in Section~\ref{ssec:sum_factorization_algo} and then consumed by the forward $Z$ contraction, so the forward pass does not materialize a full-volume reference-gradient intermediate after each one-dimensional contraction. A full-volume reference-gradient working array scales as $8 d^2 (p+2)^3$~bytes (about $72$~KB at $p{=}8$) and exceeds L1d while remaining L2-resident. Our layout carries only the $Y$-contracted intermediate across $iz$ slices, then performs one $Z$ pass over the accumulated slices. The cross-slice buffer is $3 (p+1)(p+2)^2 \cdot 8$~bytes per element (Table~\ref{tab:b1_buffers}); each channel fits within L1d on AMD~EPYC~7713 and Kunpeng~920, while the full per-element working set is L2-resident.

The on-element memory traffic in this part of the kernel is therefore dominated by L1-cache-resident accesses to \texttt{B}, \texttt{G}, \texttt{sm0}, and \texttt{sm1}; the dominant streaming reads inside the element loop are the input slice $\mathtt{x}(\cdot,\cdot,iz,c,e)$ and the per-quadrature-point material parameters $\lambda(\alpha,e), \mu(\alpha,e)$. The full three-dimensional basis-gradient matrix $G_{\mathrm{3D}}$ of the dense formulation, whose size is $O((p+1)^{2d})$ per element, never enters the working set.

\paragraph{Pointwise stress with structured Voigt arithmetic.}
At each quadrature point $\alpha = (qx,qy,qz)$ we compute the symmetric Cauchy stress $\bm{\sigma}$ from the gradient by evaluating Hooke's law (Equation~\eqref{eq:hookes_law}) with the per-quadrature-point material parameters $\lambda(\alpha,e)$ and $\mu(\alpha,e)$, weighted by $w_\alpha \det(J)$. We store $\bm{\sigma}$ in the six-component Voigt buffer \texttt{stress[6]\allowbreak[Q1D]\allowbreak[Q1D]\allowbreak[Q1D]} using the zero-based order $[00,11,22,01,02,12]$. The Voigt constitutive form shown in Equation~\eqref{eq:mult} is \emph{not} implemented as a dense matrix--vector product; instead the kernel exploits the block structure of the matrix to read each material coefficient once and reuse it across the three diagonal entries:
\begin{align*}
\sigma_{00} &= \lambda_w \,\nabla\!\cdot\!u + 2\mu_w\,\partial_x u_x, \quad
\sigma_{11} = \lambda_w \,\nabla\!\cdot\!u + 2\mu_w\,\partial_y u_y, \quad
\sigma_{22} = \lambda_w \,\nabla\!\cdot\!u + 2\mu_w\,\partial_z u_z,\\
\sigma_{01} &= \mu_w(\partial_x u_y + \partial_y u_x), \quad
\sigma_{02} = \mu_w(\partial_x u_z + \partial_z u_x), \quad
\sigma_{12} = \mu_w(\partial_y u_z + \partial_z u_y),
\end{align*}
where $\lambda_w = w_\alpha\det(J)\,\lambda(\alpha,e)$ and $\mu_w = w_\alpha\det(J)\,\mu(\alpha,e)$, and $\nabla\!\cdot\!u = \partial_x u_x + \partial_y u_y + \partial_z u_z$ is reused across the three diagonal entries. This is the structured ``Voigt arithmetic'' introduced in Section~\ref{ssec:voigt_optimization}.

\paragraph{Backward sum factorization.}
The operator action $\mathbf{y} += \int_\Omega \bm{\sigma}(\mathbf{u}) : \nabla \mathbf{v}\, d\Omega$ is implemented as the transpose of the forward pass. For each output component $q\in\{0,1,2\}$, the kernel first reconstructs the $q$-th row of $\bm{\sigma} J^{-T}$ at every quadrature point from the symmetric Voigt buffer. The reconstruction uses the symmetry $\sigma_{ij} = \sigma_{ji}$ (so $\sigma_{10}$ reads the same memory cell as $\sigma_{01}$) and the precomputed $J^{-1}$:
\[
Q_{q,m,\alpha} \;=\; \sum_{i=0}^{2} \sigma_{q,i,\alpha}\, (J^{-1})_{m,i},
\]
where $\sigma_{q,i,\alpha}$ is taken from the appropriate cell of the Voigt buffer following the index map $\sigma_{00}\to\!0$, $\sigma_{11}\to\!1$, $\sigma_{22}\to\!2$, $\sigma_{01}\to\!3$, $\sigma_{02}\to\!4$, $\sigma_{12}\to\!5$. The result is stored in a per-output-component buffer \texttt{Q\_vec\_q[3][Q1D][Q1D][Q1D]}.

The contraction back to nodal values is then performed in three one-dimensional sweeps, mirroring the forward pass:
\begin{enumerate}
\item \emph{$Z$ contraction.} For each fixed output $iz \in \{0,\dots,p\}$, contract $Q_{q,m,\alpha}$ along $qz$ against either $G(iz,qz)$ if $m=2$ or $B(iz,qz)$ otherwise, writing to a small persistent buffer \texttt{tmpZ[3][Q1D][Q1D]}.
\item \emph{$Y$ contraction.} Contract \texttt{tmpZ} along $qy$ against $G(iy,qy)$ if $m=1$ or $B(iy,qy)$ otherwise, writing to \texttt{tmpY} with shape $3\times\mathtt{D1D}\times\mathtt{Q1D}$.
\item \emph{$X$ contraction and accumulation.} Contract \texttt{tmpY} along $qx$ against $G(ix,qx)$ if $m=0$ or $B(ix,qx)$ otherwise, sum the three $m$-channels, and accumulate the scalar into \texttt{y(ix, iy, iz, q, e)}.
\end{enumerate}
The choice of $G$ versus $B$ along each direction encodes the reference-space derivative $\partial_m$ on the test-function side: applying $G$ in direction $m$ and $B$ in the other two reproduces $\partial_m$, and summing the three $m$-channels realizes the divergence-type contraction $\sum_m \partial_m(\cdot)$ without ever forming the full $G_{3D}$ matrix.

\begin{figure}[htbp]
    \centering
    \includegraphics[width=1\linewidth]{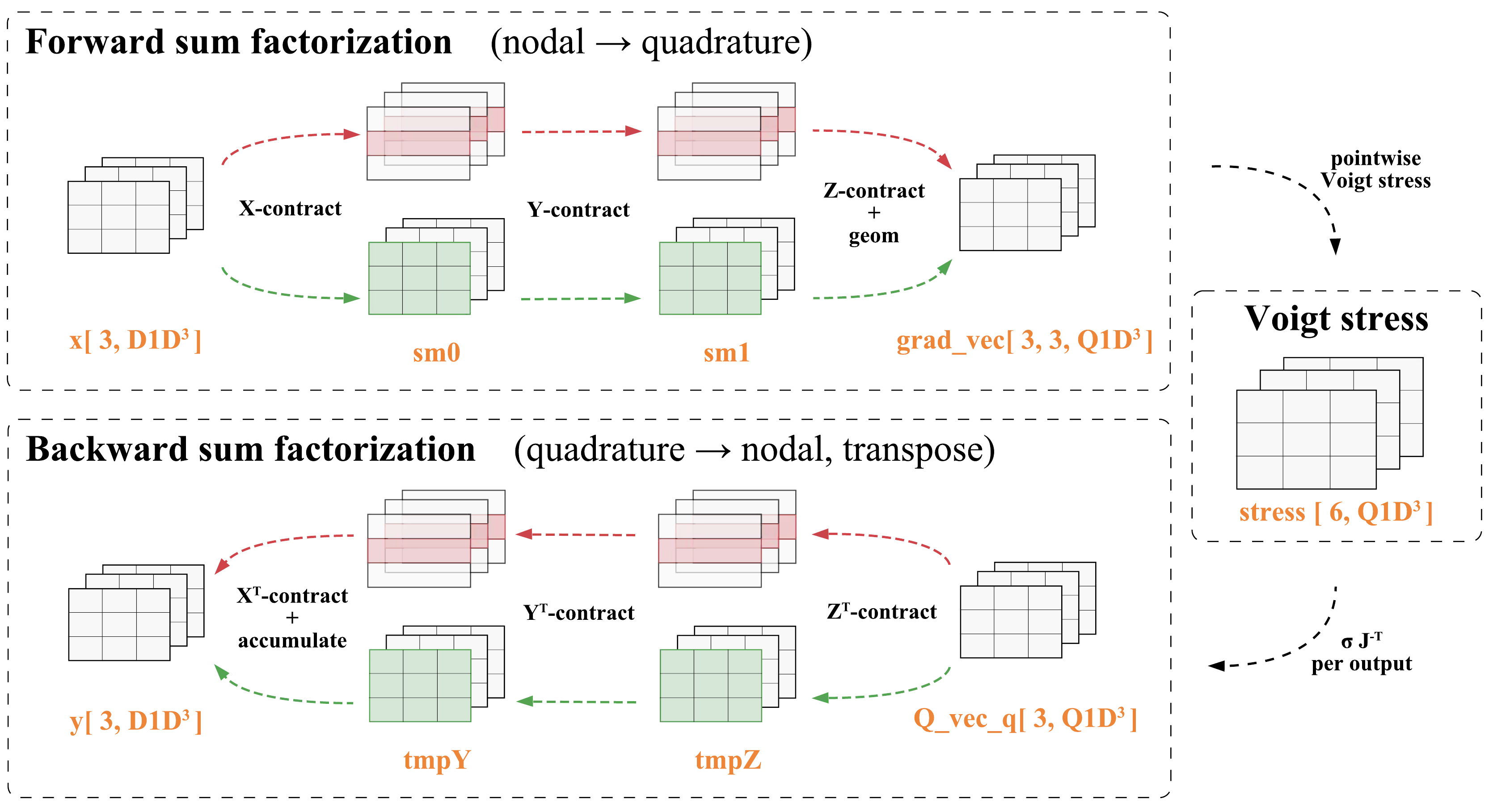}
    \caption{Element-local dataflow of PAop. The upper row shows the forward nodal-to-quadrature contractions and pointwise stress evaluation; the lower row shows the transpose contractions back to nodal values. Dashed paths contrast a direct line-wise handoff between one-dimensional contractions and the slice-wise handoff used in PAop.}
    \Description{Dataflow diagram showing forward tensor contractions, pointwise stress evaluation, backward contractions, and the slice-wise handoff used by PAop.}
    \label{fig:dataflow_pattern}
\end{figure}

\paragraph{Buffers, lifetimes, and per-element memory.}
Table~\ref{tab:b1_buffers} summarizes the on-chip buffers used by the kernel along with their symbolic size, an estimate at $p=8$, and their lifetime; see Figure~\ref{fig:dataflow_pattern} for the fiber-wise and slice-wise handoff patterns these buffers serve.

\begin{table}[htbp]
\centering
\caption{On-chip buffers in \texttt{My3DAddMultPA\_<D1D,Q1D>} for $d=3$ linear elasticity. Sizes are reported in 8-byte doubles; the symbolic-size column uses $(p+1)$ for the 1D DoFs and $(p+2)$ for the 1D quadrature points (corresponding to the source-side template parameters \texttt{D1D} and \texttt{Q1D} introduced in Section~\ref{ssec:sum_factorization_algo}). In the buffer names, \texttt{MD1} and \texttt{MQ1} are the compile-time maximum dimensions associated with \texttt{D1D} and \texttt{Q1D}.}
\label{tab:b1_buffers}
\resizebox{\columnwidth}{!}{%
\begin{tabular}{llll @{\quad} llll}
\toprule
\textbf{Buffer} & \textbf{Sym.\ size} & \textbf{KB} & \textbf{Lifetime} & \textbf{Buffer} & \textbf{Sym.\ size} & \textbf{KB} & \textbf{Lifetime} \\
\midrule
\texttt{BG[2][MQ1*MD1]}        & $2(p+1)(p+2)$     & $1.4$  & per-element      & \texttt{stress[6][.]}      & $6(p+2)^3$    & $48.0$ & per-element      \\
\texttt{sm0[2][.]}             & $2(p+1)(p+2)$     & $1.4$  & per-$iz$ slice   & \texttt{Q\_vec\_q[3][.]}   & $3(p+2)^3$    & $24.0$ & per-output-$q$   \\
\texttt{sm1[3][.]}             & $3(p+1)(p+2)^2$   & $21.6$ & cross-$iz$ slice & \texttt{tmpZ[3][.]}        & $3(p+2)^2$    & $2.4$  & per-output-$iz$  \\
\texttt{grad\_vec[d][d][.]}    & $d^2 (p+2)^3$     & $72.0$ & per-element      & \texttt{tmpY[3][.]}        & $3(p+1)(p+2)$ & $2.2$  & per-output-$iz$  \\
\bottomrule
\end{tabular}%
}
\end{table}

The dominant on-chip footprint at $p=8$ is the full-volume \texttt{grad\_vec} ($72$~KB) and \texttt{stress} ($48$~KB); the cross-slice buffer \texttt{sm1} ($21.6$~KB) is about $2.2$--$3.3\times$ smaller than these, and the inner-pass scratchpad \texttt{sm0} ($1.4$~KB) together with the per-output slice buffers \texttt{tmpZ}/\texttt{tmpY} ($\sim 2.2$--$2.4$~KB) are an order of magnitude smaller still. With both AMD~EPYC~7713 and Kunpeng~920 providing $\geq 512$~KB of L2 per core, the entire per-element working set fits in L2 throughout $p \in \{1,\dots,8\}$, which is the prerequisite for the sustained DRAM bandwidth utilization reported in Section~\ref{ssec:flops_dof}.

\paragraph{Relation to scalar tensor-product kernels.}
The forward and backward contractions inherit the slice-wise, loop-tiled evaluation pattern of Kronbichler and Kormann~\cite{kronbichler2019fast}. PAop adapts this dataflow to vector linear elasticity by carrying the three displacement components, using six-component Voigt stress storage with structured arithmetic, and applying the affine geometry transform inside the element-local fused kernel.

\section{Experiments}
\label{sec:experiments}

This section evaluates FA, PA, and PAop on x86 and ARM CPU platforms, covering solver time, memory footprint, kernel throughput, FLOPs/DoF, and hardware-counter behavior.

\subsection{Experimental Setup}
\label{ssec:setup}

We first define the hardware platforms, methods, metrics, and benchmark problem used in the evaluation.

\subsubsection{Hardware and Software Environment}
\label{sssec:hw_sw}

We conducted performance evaluations on two architecturally distinct platforms: AMD~EPYC~7713 (x86-64) and Huawei~Kunpeng~920 (AArch64). Detailed hardware specifications and software configurations are provided in Table~\ref{tab:cpu_specs_wide_tabularx}. We used MFEM~v4.8 as the foundational library. All experiments use a pure-MPI configuration with 64 MPI ranks per platform, one rank per physical core; on AMD~EPYC~7713 the active cores are pinned to the single-socket NUMA group, and on Kunpeng~920 they span NUMA nodes~$0$--$1$. The DRAM bandwidth values are sustained LIKWID \texttt{likwid-bench} throughputs~\cite{treibig2010likwid} measured on the same NUMA pinning used for the application runs; peak compute values are from \texttt{cpufp}~\cite{pigirons_cpufp_2022} on the same single socket.
\begin{table*}[htbp]
\centering
\caption{\small Hardware and software configuration. Bandwidth and peak compute were measured under the same pinning as the application runs; cache capacities are per-core vendor values. On both platforms the code was compiled with \texttt{-O3 -fopenmp}: AMD~EPYC~7713 on Ubuntu~20.04 with gcc~13.3.1 and \texttt{-march=znver3 -mavx2 -mfma}; Kunpeng~920 on KylinSec~OS Linux~3 with gcc for openEuler~3.0.3 and \texttt{-march=native -ffast-math}.}
\label{tab:cpu_specs_wide_tabularx}

\begin{tabularx}{\textwidth}{@{} l >{\centering\arraybackslash}X >{\centering\arraybackslash}X @{}}
\toprule

& \textbf{AMD EPYC 7713}
& \textbf{HUAWEI Kunpeng 920 V200 7270Z} \\
\midrule

cores & $2 \times 64$ & $1 \times 64$ \\
frequency (base / max) & 2.0 / 3.675\,GHz & 2.9 / 2.9\,GHz \\
cache per core (L1d / L2 / L3) & 32\,KB / 512\,KB / 4\,MB & 64\,KB / 512\,KB / 1.75\,MB \\
compute peak (SIMD) & 3027.44\,GFLOP/s (AVX2) & 2597.9\,GFLOP/s (SVE) \\
memory BW (size) & 164.66\,GB/s (256\,GB) & 180\,GB/s (512\,GB) \\
\bottomrule
\end{tabularx}

\end{table*}

\subsubsection{Methods Under Test}
\label{sssec:methods}

Within the GMG preconditioning framework of Section~\ref{sec:gmg_architecture}, we compare three operator implementations: \textbf{FA} (full assembly, MFEM \texttt{AssemblyLevel::FULL}), \textbf{PA} (the unmodified MFEM linear-elasticity \texttt{ElasticityIntegrator} at \texttt{AssemblyLevel::PARTIAL}), and \textbf{PAop} (the optimized PA operator built in Section~\ref{sec:optimization}).

\subsubsection{Performance Metrics}
\label{sssec:metrics}

We report total time, peak memory usage, throughput in MDoF/s, and computational throughput in GFLOP/s. Phase timings are benchmark-driver wall-clock timings measured with MFEM \texttt{StopWatch}. Kernel timings are accumulated inside matrix-free \texttt{AddMult} calls and reported as the maximum over MPI ranks. Hardware-counter runs use LIKWID marker regions around the same \texttt{AddMult} calls; throughput scopes are defined in Section~\ref{ssec:flops_dof}.

\subsubsection{Benchmark Problem}
\label{sssec:benchmark}

All experiments solve the static linear-elasticity problem defined by Equation~\eqref{eq:elasticity_strong} (Section~\ref{ssec:governing_equations}), with the isotropic Hooke's-law constitutive relation $\bm{\sigma}(\bm{u}) = \lambda(\nabla \cdot \bm{u})\bm{I} + 2\mu\,\bm{\varepsilon}(\bm{u})$ and small-strain tensor $\bm{\varepsilon}(\bm{u}) = \tfrac{1}{2}(\nabla\bm{u} + (\nabla\bm{u})^T)$.

\noindent\textbf{Domain and mesh.}~We use the two-material cantilever beam from MFEM Example~2 (\texttt{data/beam-hex.mesh}, an $8\times 1\times 1$ structured hexahedral block; the mesh is shipped with MFEM and serves as the reference benchmark for \texttt{ex2p}). The serial mesh is uniformly refined until the element count reaches $\sim 1000$, then partitioned across $64$ MPI ranks; additional parallel refinements bring the finest-level DoF count to one of the reported sizes ($0.84$M, $6.5$M, or $51.17$M, depending on the experiment). The fixed-DoF studies in Sections~\ref{ssec:macro_perf} and \ref{ssec:micro_perf} compensate increases in $p$ by skipping a corresponding number of $h$-refinements so that the finest-level DoF count is held approximately constant.

\noindent\textbf{Material and load.}~The two element attributes carry contrasting Lam{\'e} parameters: attribute~1 has $\lambda_1 = \mu_1 = 50$ and attribute~2 has $\lambda_2 = \mu_2 = 1$ (a $50{:}1$ stiffness contrast, matching the MFEM \texttt{ex2p} reference setup). The Dirichlet boundary $\Gamma_D$ is the boundary attribute~$1$ face (the clamped end of the beam), where homogeneous displacement is imposed. The Neumann traction $\bm{t}$ is applied on boundary attribute~$2$ as a constant downward pull $\bm{t} = (0,\,0,\,-10^{-2})$, with the remaining boundary attributes traction-free.

\noindent\textbf{Discretization and stopping criterion.}~We use $H^1$-conforming continuous Galerkin elements of polynomial degree $p \in \{1,2,4,8\}$ with $(p+1)$ tensor-product Gauss--Legendre--Lobatto nodes and the MFEM default over-integration rule with $(p+2)$ Gauss points per dimension. The discrete linear system is solved by the outer preconditioned conjugate-gradient solver, MFEM \texttt{CGSolver} with \texttt{SetRelTol(1e-6)} (implemented for preconditioned solves as $(B r_k,r_k)^{1/2}/(B r_0,r_0)^{1/2}\le10^{-6}$) and a $5{,}000$-iteration cap; iteration counts are reported in Section~\ref{ssec:preconditioner_perf}.

\noindent\textbf{Reproduction.}~Each reported configuration is identified by the tuple $(p, h, \text{solver}, k, \text{strong\_thresh})$, together with the benchmark problem, material parameters, and stopping criterion given above.

\subsection{Preconditioner Performance Comparison}
\label{ssec:preconditioner_perf}

To select a preconditioning strategy for the remainder of Section~\ref{sec:experiments}, we compare four combinations on the GMG hierarchy of Section~\ref{sec:gmg_architecture}: \textbf{FA + AMG}, \textbf{PA + Jacobi}, \textbf{FA + GMG}, and \textbf{PA + GMG}. PA + Jacobi is retained as a simple directly matrix-free baseline; AMG requires explicit matrix entries and therefore cannot serve as a fine-level PA preconditioner without assembling an auxiliary matrix. Both the four-solver comparison and the parameter sweep summarized below were measured with a dedicated, self-contained preconditioner benchmark driver built on MFEM's native \texttt{ElasticityIntegrator}; all configurations in Table~\ref{tab:gmg_cost_breakdown} therefore share one controlled measurement setup, while absolute wall-clock times are not directly comparable with those of the end-to-end solver driver used in Sections~\ref{ssec:macro_perf} and~\ref{ssec:micro_perf}.

\begin{table}[htbp]
\centering
\caption{\small GMG cost breakdown for four preconditioners on Kunpeng with 64 MPI ranks at (a) $0.84$M and (b) $6.5$M DoFs. Times are seconds; \texttt{Total} includes mesh, FE-space, and bilinear-form setup.}
\label{tab:gmg_cost_breakdown}
\setlength{\tabcolsep}{3pt}

\begin{minipage}[t]{0.49\linewidth}
\centering
\textbf{(a) $0.84$M DoFs}\\[0.3em]
\resizebox{\linewidth}{!}{%
\begin{tabular}{
    @{}
    c l
    S[table-format=4.0]
    S[table-format=4.2]
    S[table-format=4.2]
    S[table-format=4.2]
    S[table-format=4.2]
    @{}
}
\toprule
$p$ & {Solver} & {Iters} & {Prec.} & {Form-LS} & {Solve} & {Total} \\
\midrule
\multirow{4}{*}{1}
& \texttt{fa\_amg} &   65 &   0.15 & 0.13 &    2.16 &    2.50 \\
& \texttt{fa\_gmg} &    6 &   0.35 & 0.00 &    0.16 &    0.57 \\
& \texttt{pa\_jac} &  915 &   0.08 & 0.03 &   35.89 &   36.07 \\
& \texttt{pa\_gmg} &    6 &   0.52 & 0.03 &    1.47 &    2.08 \\
\midrule
\multirow{4}{*}{2}
& \texttt{fa\_amg} &   47 &   0.35 & 0.26 &    3.45 &    4.07 \\
& \texttt{fa\_gmg} &    7 &   0.71 & 0.00 &    0.32 &    1.05 \\
& \texttt{pa\_jac} & 1195 &   0.04 & 0.02 &   19.41 &   19.49 \\
& \texttt{pa\_gmg} &    7 &   0.27 & 0.02 &    0.85 &    1.15 \\
\midrule
\multirow{4}{*}{4}
& \texttt{fa\_amg} &   52 &   3.53 & 0.89 &   13.50 &   17.92 \\
& \texttt{fa\_gmg} &    9 &   4.88 & 0.00 &    1.20 &    6.09 \\
& \texttt{pa\_jac} & 1548 &   0.07 & 0.02 &   34.37 &   34.47 \\
& \texttt{pa\_gmg} &    9 &   0.32 & 0.02 &    1.23 &    1.58 \\
\midrule
\multirow{4}{*}{8}
& \texttt{fa\_amg} &   57 & 457.50 & 5.48 &   60.29 &  523.32 \\
& \texttt{fa\_gmg} &   12 & 464.99 & 0.01 &    6.82 &  471.87 \\
& \texttt{pa\_jac} & 1937 &   0.27 & 0.07 &  127.27 &  127.66 \\
& \texttt{pa\_gmg} &   12 &   0.97 & 0.07 &    4.49 &    5.58 \\
\bottomrule
\end{tabular}%
}
\end{minipage}%
\hfill
\begin{minipage}[t]{0.49\linewidth}
\centering
\textbf{(b) $6.5$M DoFs}\\[0.3em]
\resizebox{\linewidth}{!}{%
\begin{tabular}{
    @{}
    c l
    S[table-format=4.0]
    S[table-format=4.2]
    S[table-format=4.2]
    S[table-format=4.2]
    S[table-format=4.2]
    @{}
}
\toprule
$p$ & {Solver} & {Iters} & {Prec.} & {Form-LS} & {Solve} & {Total} \\
\midrule
\multirow{4}{*}{1}
& \texttt{fa\_amg} &   63 &   1.22 & 1.58 &   19.17 &   22.47 \\
& \texttt{fa\_gmg} &    7 &   3.39 & 0.00 &    1.34 &    5.23 \\
& \texttt{pa\_jac} & 1829 &   0.67 & 0.31 &  564.25 &  565.75 \\
& \texttt{pa\_gmg} &    7 &   4.22 & 0.28 &   13.10 &   18.09 \\
\midrule
\multirow{4}{*}{2}
& \texttt{fa\_amg} &   60 &   2.78 & 2.52 &   35.74 &   41.13 \\
& \texttt{fa\_gmg} &    7 &   6.21 & 0.00 &    2.42 &    8.72 \\
& \texttt{pa\_jac} & 2387 &   0.35 & 0.13 &  336.77 &  337.33 \\
& \texttt{pa\_gmg} &    7 &   2.26 & 0.13 &    6.76 &    9.24 \\
\midrule
\multirow{4}{*}{4}
& \texttt{fa\_amg} &   56 &  29.76 & 5.13 &  116.09 &  151.01 \\
& \texttt{fa\_gmg} &    8 &  39.14 & 0.01 &    8.54 &   47.73 \\
& \texttt{pa\_jac} & 3097 &   0.50 & 0.18 &  551.35 &  552.06 \\
& \texttt{pa\_gmg} &    8 &   2.52 & 0.17 &    8.72 &   11.45 \\
\midrule
\multirow{4}{*}{8}
& \texttt{fa\_amg} &   61 &3652.73 &58.52 & 1122.78 & 4834.10 \\
& \texttt{fa\_gmg} &   12 &3734.54 & 0.05 &   53.79 & 3788.44 \\
& \texttt{pa\_jac} & 3864 &   1.63 & 0.52 & 2031.02 & 2033.24 \\
& \texttt{pa\_gmg} &   12 &   7.19 & 0.52 &   35.69 &   43.47 \\
\bottomrule
\end{tabular}%
}
\end{minipage}

\end{table}

Table~\ref{tab:gmg_cost_breakdown} presents the per-phase cost breakdown of the four solver variants on Kunpeng with 64 MPI ranks, split into the two DoF scales (a) $0.84$M and (b) $6.5$M. The three reported phases are \texttt{Prec.} (preconditioner setup, including the AMG hierarchy build for \texttt{fa\_amg} and the level-smoother plus coarse-AMG setup for \texttt{*\_gmg}), \texttt{Form-LS} (\texttt{FormSystemMatrix} / \texttt{FormFineLinearSystem}, including the application of essential Dirichlet conditions), and \texttt{Solve} (outer PCG iterations including all $V$-cycles, smoothings and the inner coarse PCG). \texttt{Total} additionally includes mesh + FE-space construction ($\le 0.50$~s in all configurations) and bilinear-form assembly ($\le 0.01$~s), which are not shown as separate columns.

Three effects are visible. First, the \texttt{Solve} column is dominated by the iteration count, which drops by $\sim 150\text{--}390\times$ from Jacobi to GMG; combined with the $\sim 5.7\text{--}7.5\times$ higher per-iteration cost of a GMG $V$-cycle, this yields a $\sim 23\text{--}63\times$ \texttt{Solve}-time reduction. Second, the \texttt{Prec.} column is dominated by AMG setup, which grows sharply for $p=8$ on \texttt{fa\_amg} and \texttt{fa\_gmg} ($\sim 458$--$465$~s on 0.84M DoFs and $\sim 3653$--$3735$~s on 6.5M DoFs) because the assembled high-order stiffness matrix has many more nonzero entries per row, inflating AMG coarsening cost. Third, for \texttt{pa\_gmg}, the \texttt{Prec.} column stays under $8$~s in all cases because AMG only sees the coarsest-level matrix, so \texttt{pa\_gmg} \texttt{Total} remains competitive across all $p$ while \texttt{fa\_amg} and \texttt{fa\_gmg} diverge at $p=8$. We therefore use GMG as the unified preconditioner for the remainder of Section~\ref{sec:experiments}.

A small parameter sweep at $p=4$ varied Chebyshev degree $k\in\{1,2,3\}$ and BoomerAMG systems/elasticity options, including strong-threshold settings. The GMG variants were insensitive to strong-threshold tuning because AMG acts only on the coarsest matrix, while $k=2$ gave the best or near-best total time among the tested smoother degrees. We therefore use Chebyshev degree $k=2$ and the default BoomerAMG systems configuration described in Section~\ref{sec:gmg_architecture}.

\subsection{Solver-Level Performance Analysis}
\label{ssec:macro_perf}

Having fixed GMG as the unified preconditioner, this section reports the end-to-end performance of PAop on a large-DoF elasticity problem and compares it to the FA and PA baselines defined in Section~\ref{sssec:methods}.
\begin{figure}[htbp]
    \centering
    \includegraphics[width=\textwidth]{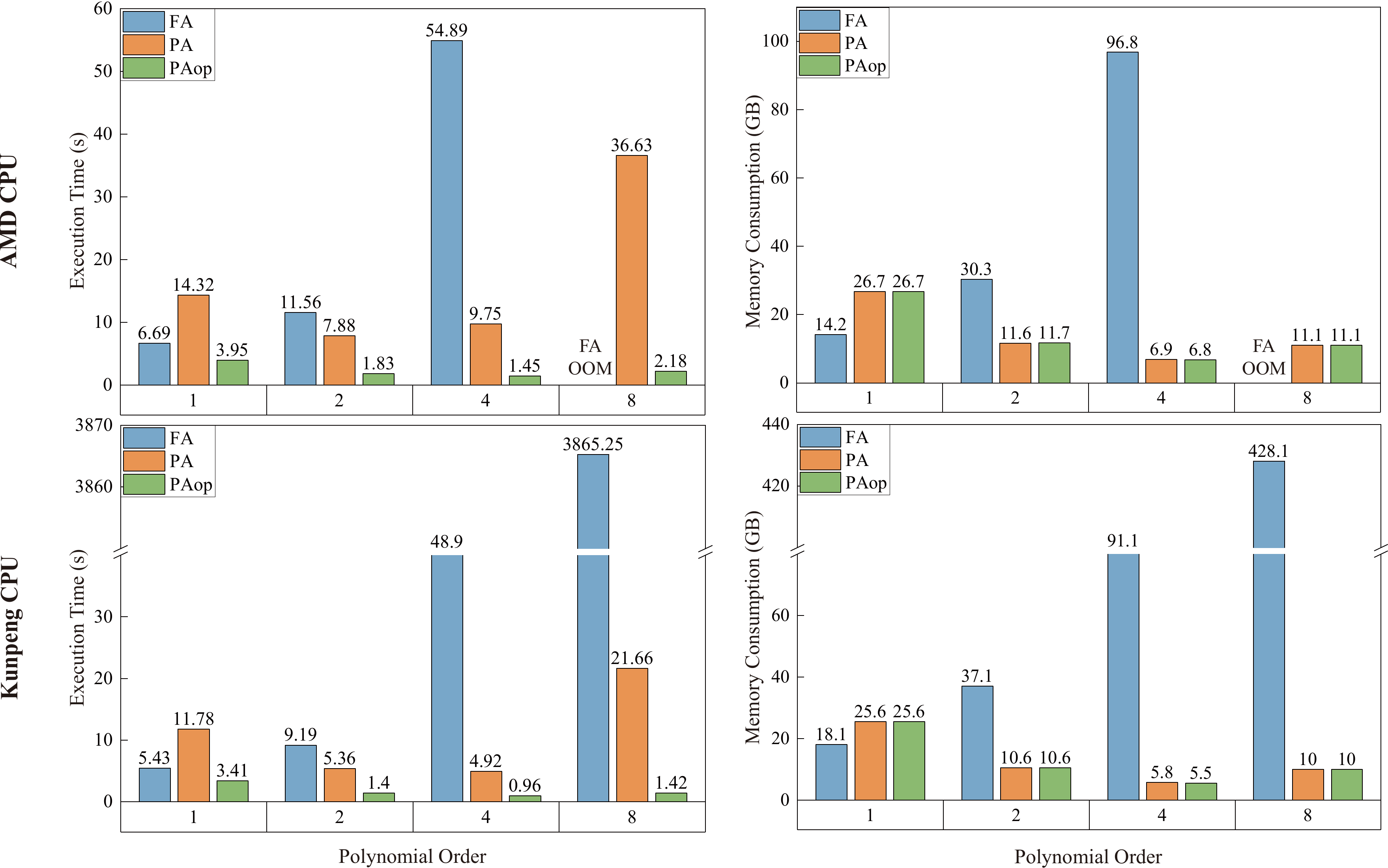}
    \caption{End-to-end wall-clock time and peak memory usage for Full Assembly (FA), baseline Partial Assembly (PA), and the optimized operator (PAop) on AMD~EPYC~7713 and Kunpeng~920 with 64 MPI ranks at approximately $6.5$~million finest-level vector DoFs across $p \in \{1,2,4,8\}$. All methods use the same GMG preconditioner and the same PCG stopping criterion defined in Section~\ref{sssec:benchmark}.}
    \Description{Bar charts comparing solve time and peak memory for FA, PA, and PAop on AMD and Kunpeng across polynomial orders.}
    \label{fig:placeholder_exp1}
\end{figure}

At a fixed problem size of approximately $6.5$~million DoFs, we compare the total solve time and peak memory usage of FA, PA, and PAop across $p \in \{1,2,4,8\}$ (Figure~\ref{fig:placeholder_exp1}). Under the same GMG preconditioner and stopping criterion (Section~\ref{sssec:benchmark}), the iteration count is identical across the three variants at each polynomial degree; therefore, the solver-level timing differences mainly reflect differences in operator-application cost rather than convergence behavior. PAop is up to $16.8\times$ faster than PA at $p=8$ and attains the smallest end-to-end solve time at $p=4$. This solver-level minimum is distinct from the operator-throughput sweet spot: Figure~\ref{fig:placeholder_mdof} isolates kernel-time throughput and shows the optimized operator peaking around $p=6$ while remaining efficient at $p=8$. The peak memory of FA grows with $p$ while the footprint of PA and PAop is approximately flat, reflecting the well-known matrix-free advantage in the high-order regime.

The $51.17$M-DoF test in Table~\ref{tab:macro_perf_51M} reproduces the same trend at larger scale on the Kunpeng platform. FA hits an out-of-memory error for $p \ge 4$ on the $512$~GB-per-node test machine, and the PAop end-to-end speedup over PA reaches $15.56\times$ at $p=8$. The steep growth of the PA \texttt{Solve} column from $p=4$ to $p=8$ reflects the $O((p+1)^6)$ contraction cost of the unmodified MFEM linear-elasticity PA kernel (Section~\ref{ssec:baseline_analysis}).

\begin{table*}[htbp]
\centering
\caption{\small Solver-level performance comparison on the 51.17M-DoF problem on Kunpeng~920 with 64 MPI ranks. All methods use the GMG preconditioner of Section~\ref{sec:gmg_architecture} and the stopping criterion of Section~\ref{sssec:benchmark}. \texttt{Assembly} combines GMG/preconditioner setup and \texttt{FormFineLinearSystem} time; \texttt{Solve} is the outer GMG-preconditioned PCG solve; \texttt{Total} is end-to-end time-to-solution. Peak memory is reported in GB.}
\label{tab:macro_perf_51M}

\sisetup{
    table-align-text-post=false,
    output-decimal-marker={.}
}
\begin{tabular}{
    @{}
    c l S[table-format=2.0]
    S[table-format=2.2] S[table-format=3.2] S[table-format=3.2]
    r
    r
    S[table-format=3.2]
    r
    @{}
}
\toprule
\textbf{Order} & \textbf{Algorithm} & {\textbf{Iters}} & {\textbf{Assembly}} & {\textbf{Solve}} & {\textbf{Total}} & {\textbf{Speedup}} & {\textbf{Speedup}} & {\textbf{Peak Mem.}} & {\textbf{FA Mem. /}} \\
\textbf{(p)} & & & {\textbf{(s)}} & {\textbf{(s)}} & {\textbf{(s)}} & {\textbf{(vs. FA)}} & {\textbf{(vs. PA)}} & {\textbf{(GB)}} & {\textbf{Method Mem.}} \\
\midrule
\multirow{3}{*}{1}
 & FA                 & 6 & 45.12 &  9.47 & 59.78 & $1.00\times$  & {\textemdash} & 141.36 & $1.00\times$  \\
 & PA                 & 6 & 21.98 & 59.66 & 88.30 & {\textemdash} & $1.00\times$  & 196.03 & {\textemdash} \\
 & PAop               & 6 &  6.87 & 14.04 & 26.28 & $2.28\times$  & $3.36\times$  & 196.01 & $0.72\times$  \\
\midrule
\multirow{3}{*}{2}
 & FA                 & 7 & 93.43 & 24.33 & 118.58& $1.00\times$  & {\textemdash} & 321.56& $1.00\times$  \\
 & PA                 & 7 & 10.01 & 31.51 & 42.94 & {\textemdash} & $1.00\times$  & 78.55 & {\textemdash} \\
 & PAop               & 7 &  3.10 &  8.06 & 12.10 & $9.80\times$  & $3.55\times$  & 78.55 & $4.09\times$  \\
\midrule
\multirow{3}{*}{4}
 & FA                 & \multicolumn{8}{c}{OOM (Out of Memory)} \\
 & PA                 & 8 &  8.74 & 30.60 & 40.24 & {\textemdash} & $1.00\times$  & 39.82 & {\textemdash} \\
 & PAop               & 8 &  1.79 &  6.19 &  8.38 & {\textendash} & $4.80\times$  & 39.82 & {\textendash} \\
\midrule
\multirow{3}{*}{8}
 & FA                 & \multicolumn{8}{c}{OOM (Out of Memory)} \\
 & PA                 & 11 & 27.06 & 131.56 & 161.00 & {\textemdash} & $1.00\times$  & 33.45 & {\textemdash} \\
 & PAop               & 11 &  1.54 &  8.43 & 10.35 & {\textendash} & $15.56\times$ & 33.44 & {\textendash} \\
\multicolumn{10}{l}{\footnotesize\textit{Note:} The memory ratio is FA peak memory divided by method peak memory; values below $1$ mean method memory exceeds FA.} \\
\multicolumn{10}{l}{\footnotesize At $p \ge 4$, FA exceeded the $512$~GB-per-node memory budget; FA-relative speedup and memory ratios are therefore undefined.} \\
\bottomrule
\end{tabular}
\end{table*}

\subsection{Operator-Level Performance Analysis}
\label{ssec:micro_perf}

This section analyzes the matrix-free \texttt{AddMult} kernel, comparing PAop with the MFEM PA baseline.

\subsubsection{Throughput definitions, FLOPs/DoF, and hardware-counter measurements}
\label{ssec:flops_dof}

Throughput in MDoF/s can be measured at solver, kernel-time, or operator-implied scope; this work distinguishes all three because they answer different performance questions.

\begin{figure}[htbp]
    \centering
    \includegraphics[width=\textwidth]{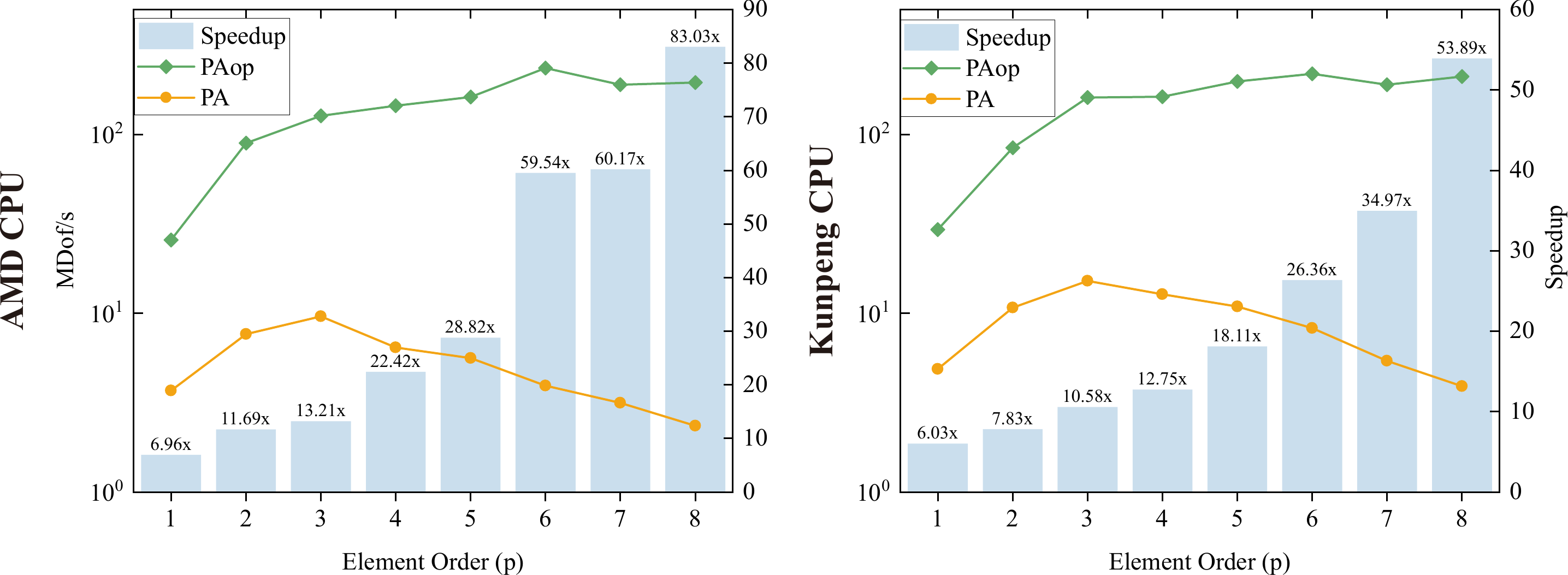}
    \caption{Kernel-time throughput (MDoF/s, $T_\text{L2}$ in Section~\ref{ssec:flops_dof}) and PAop/PA speedup ratio of the \texttt{AddMult} function on AMD (left) and Kunpeng (right) at $\sim 6.5$M DoFs with 64 MPI ranks across $p=1,\ldots,8$; interpretation in Section~\ref{ssec:flops_dof}.}
    \Description{Line plots of kernel-time throughput and PAop-over-PA speedup on AMD and Kunpeng as polynomial order increases.}
    \label{fig:placeholder_mdof}
\end{figure}

\paragraph{Throughput scopes used in this work.}
\begin{enumerate}
\item[L1] \emph{Solver throughput.} $T_\text{L1} = (\text{iterations} \cdot \text{DoFs}) / t_\text{solve}$, where $t_\text{solve}$ is the wall-clock time of the GMG-preconditioned PCG including the coarse $V$-cycle, all transfers, and the smoother. This is the metric a user observes end-to-end.
\item[L2] \emph{Kernel-time throughput.} $T_\text{L2} = (\text{iterations} \cdot \text{DoFs}) / t_\text{kernel}$, where $t_\text{kernel}$ is the total wall-clock time spent inside the matrix-free operator, summed across all PCG iterations and across every GMG level at which the operator is invoked (the fine-level CG matrix-vector application plus all pre- and post-smoother applications on every matrix-free level of the multigrid hierarchy). This is the metric reported in Figure~\ref{fig:placeholder_mdof}.
\item[L3] \emph{Kernel-implied (operator-level) throughput.} $T_\text{L3} = G \cdot 10^9 \cdot \text{DoFs} / (\text{FLOPs/elem} \cdot n_\text{elem})$, where $G$ is the sustained floating-point rate (in GFLOP/s) measured by LIKWID inside the kernel and FLOPs/elem is the analytical operation count of the operator. This is the metric reported by Schussnig et al.\ \cite{schussnig2025matrix} and Davydov et al.\ \cite{davydov2020matrix}.
\end{enumerate}

Figure~\ref{fig:placeholder_mdof} reports $T_\text{L2}$; the cross-paper comparison in Section~\ref{ssec:flops_dof_compare} uses $T_\text{L3}$ to match the operator-level metric of prior comparators \cite{schussnig2025matrix,davydov2020matrix}.

\paragraph{FLOPs/DoF analysis.}
FLOP counts are derived from the PAop and MFEM source and summarized in Table~\ref{tab:flops_dof}. The resulting FLOPs/DoF $= F_\text{PAop}(p) / (3p^3)$ is reported with analytical and LIKWID-measured operational intensity. The denominator uses the large structured $C^0$-mesh convention that one hexahedral element contributes approximately $p^3$ scalar global DoFs; element-local storage still uses $(p+1)^3$ nodal values.

\begin{table}[htbp]
\centering
\caption{Per-element FLOP count, FLOPs per asymptotic global DoF, theoretical operational intensity, and LIKWID-measured operational intensity (RETIRED\_SSE\_AVX\_FLOPS\_ALL over total DRAM bytes) for the PAop kernel and the MFEM~v4.8 baseline at $51.17$M~DoFs on AMD EPYC~7713 with 64 MPI ranks. Here $D1D = p+1$ and $Q1D = p+2$. The FLOPs-per-element ratio (last column) grows as $O(p^2)$, which explains the migration of the operator-throughput sweet spot to high $p$.}
\label{tab:flops_dof}
\setlength{\tabcolsep}{4pt}
\begin{tabular}{c c c S[table-format=8.0] S[table-format=4.0] S[table-format=2.1] S[table-format=2.2] S[table-format=2.0]}
\toprule
$p$ & $D1D$ & $Q1D$ & {FLOPs/elem (PAop)} & {FLOPs/DoF (PAop)} & {OI (theory)} & {OI (LIKWID)} & {Ratio (Base/PAop)} \\
\midrule
1 & 2 & 3 & 7107      & 2369 & 6.6  & 4.30 & 2  \\
2 & 3 & 4 & 22892     & 954  & 7.5  & 5.72 & 2  \\
4 & 5 & 6 & 119688    & 623  & 9.6  & 6.98 & 5  \\
8 & 9 & 10 & 956048   & 622  & 13.9 & 9.34 & 14 \\
\bottomrule
\end{tabular}
\end{table}

The right-most ratio column quantifies the operator-throughput sweet-spot shift: PAop reduces the per-element FLOP count by $14\times$ relative to MFEM's baseline elasticity PA at $p=8$, against $5\times$ at $p=4$ and $2\times$ at $p=2$. Within $p \in [1,8]$, lower-order terms keep PAop FLOPs/DoF nearly constant, explaining the high-$p$ throughput plateau in Figure~\ref{fig:placeholder_mdof}.

\paragraph{Solver throughput on the AMD platform.}
At $51.17$M~DoFs on AMD~EPYC~7713 with 64~MPI~ranks, the user-facing solver throughput $T_\text{L1}$ (Section~\ref{ssec:flops_dof}) reaches $25.7 / 44.9 / 50.1$~MDoF/s at $p = 2 / 4 / 8$.

Per-$p$ LIKWID measurements at $51.17$M~DoFs provide the sustained floating-point rate, DRAM-bandwidth, and OI data for the right panel of Figure~\ref{fig:placeholder_roofline}; the OI values use the finite-mesh FLOPs/DoF accounting in Table~\ref{tab:flops_dof}.

We now examine the cross-platform throughput trajectories on both AMD and Kunpeng across $p=1,\ldots,8$. Figure~\ref{fig:placeholder_mdof} reports the kernel-time throughput (Section~\ref{ssec:flops_dof}, L2) and the PAop/PA ratio of the \texttt{AddMult} function at $\sim 6.5$M~DoFs on the two platforms. Two trajectories are visible. The baseline PA kernel throughput drops above $p=2$ as the per-element working set exceeds L1/L2 cache capacity (Section~\ref{ssec:loop_optimization}); PAop sustains throughput across the tested orders by reducing the working-set size and improving locality, peaking at $p=6$ and remaining high at $p=8$. The maximum kernel speedups are $54\times$ on Kunpeng and $83\times$ on AMD at $p=8$.

\subsubsection{Comparison with prior matrix-free elasticity codes}
\label{ssec:flops_dof_compare}

Table~\ref{tab:lit_comparison} places PAop in context with published CPU matrix-free elasticity operators from Davydov et al.\ \cite{davydov2020matrix} (deal.II, finite-strain hyperelasticity) and Schussnig et al.\ \cite{schussnig2025matrix} (ExaDG, finite-strain hyperelasticity and linear elasticity). The table reports FLOPs/DoF where derivable, operator-implied rate, and DRAM-bandwidth utilization as a normalized bandwidth-efficiency indicator; the comparison is cross-code context rather than a controlled benchmark.

\begin{table}[htbp]
\centering
\caption{\small Cross-code context for matrix-free elasticity operators on bandwidth-bound CPU configurations. Sources: this work (Section~\ref{ssec:flops_dof}), Davydov \cite[Table~2 and Table~4]{davydov2020matrix}, and Schussnig \cite[Figure~4 and Table~4]{schussnig2025matrix}. Derived entries are marked in the notes.}
\label{tab:lit_comparison}
\setlength{\tabcolsep}{3pt}
\renewcommand{\arraystretch}{1.2}
\begin{threeparttable}
\resizebox{\textwidth}{!}{%
\begin{tabular}{@{} l l c c c c @{}}
\toprule
Work & Operator & Degree $p$ & FLOPs/DoF & Rate (GDoF/s)\tnote{d} & DRAM BW utilization \\
\midrule
This work & linear elasticity PAop & $1 / 2 / 4 / 8$
  & $2369 / 954 / 623 / 622$
  & $0.24 / 0.86 / 1.24 / 1.40$
  & $78.8 / 85.9 / 66.3 / 55.6\%$ \\
Davydov 2020\tnote{a} & finite-strain tangent & $2 / 4$
  & $\approx 334 / 234$
  & $\approx 0.337 / 0.553$
  & $\approx 83\text{--}100\%$ \\
Schussnig 2025\tnote{b} & hyperelastic tangent & $4\text{--}6$
  & n.r.
  & $\approx 1.75 / 1.25$
  & $\approx 79\text{--}100\%$ \\
Schussnig 2025\tnote{c} & linear elasticity & $1\text{--}6$
  & n.r.
  & $\approx 3.3\text{--}3.7$
  & $\approx 40\text{--}77\%$ \\
\bottomrule
\end{tabular}%
}
\begin{tablenotes}\footnotesize
\item[a] Derived from Davydov 2020 Tables~2 and~4 using the reported DoF count, GFLOP/s, wall time, and bandwidth range.
\item[b] Peak operator rates and bandwidth ranges from Schussnig 2025 Figure~4 and Section~7.2; FLOP counts are not reported.
\item[c] Derived from Schussnig 2025 Table~4 relative linear-elasticity throughput and memory-transfer ratios against the fiber-model baseline.
\item[d] Raw single-application operator rate; this rate is platform-dependent and is read together with DRAM-bandwidth utilization.
\end{tablenotes}
\end{threeparttable}
\end{table}

PAop attains $55.6\%\text{--}85.9\%$ of sustained DRAM bandwidth on AMD across $p \in \{1,2,4,8\}$. The corresponding range derived from Schussnig's linear-elasticity comparison is $\approx 40\%\text{--}77\%$ \cite[Table~4]{schussnig2025matrix}. Both linear-elasticity kernels operate below the $\sim 100\%$ DRAM saturation reported for arithmetically heavier material models: with less arithmetic per loaded byte, a larger share of the runtime is exposed to latency and instruction-throughput effects rather than pure streaming, so sustained DRAM utilization stays below saturation in both codes.

\subsubsection{Roofline Model-Based Bottleneck Analysis}
\label{sssec:roofline}

\begin{figure}[htbp]
    \centering
    \includegraphics[width=\textwidth]{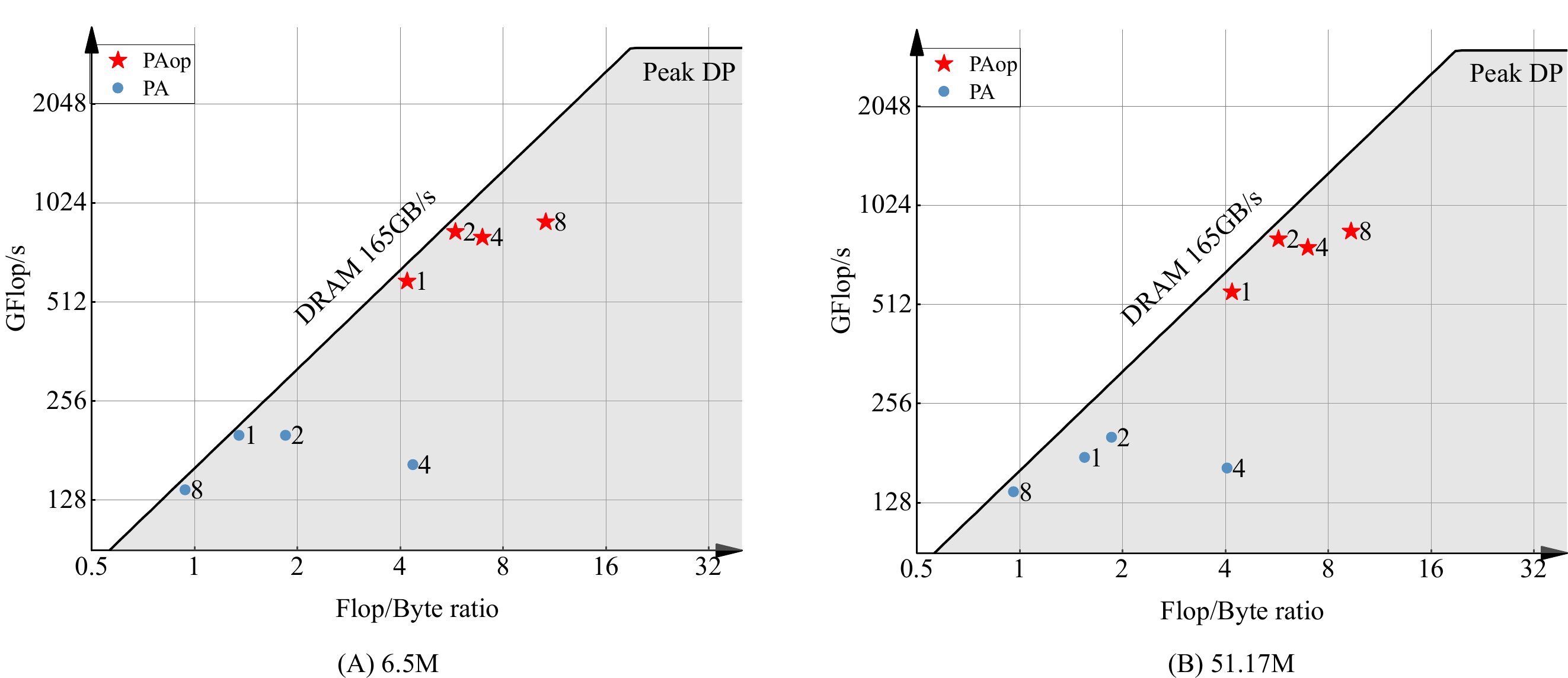}
    \caption{Roofline analysis on AMD~EPYC~7713 (64~MPI~ranks) at $6.5$M and $51.17$M DoFs. Points are measured with LIKWID; ceilings are the measured DRAM bandwidth ($164.66$~GB/s) and compute peak ($3027.44$~GFLOP/s) from Section~\ref{sssec:hw_sw}.}
    \Description{Roofline plot comparing PA and PAop operational intensity and sustained performance at two problem sizes on AMD EPYC 7713.}
    \label{fig:placeholder_roofline}
\end{figure}

Figure~\ref{fig:placeholder_roofline} plots Roofline models~\cite{williams2009roofline} for AMD~EPYC~7713 at two problem sizes ($6.5$M and $51.17$M DoFs). The full PAop optimization stack moves the data points up and to the right at both scales: PA points cluster at low operational intensity (OI $0.94$--$4.36$~FLOP/Byte), while PAop reaches OI $4.19$--$10.67$~FLOP/Byte. All PAop points lie to the left of the ridge point (OI $< 18.4$~FLOP/Byte) and remain nominally bandwidth-bound; the high-$p$ points move toward the ridge, reflecting cache residency and growing arithmetic density rather than a reversal of the optimization trend. Across the near $8\times$ change in DoF count, the relative ordering between PA and PAop is preserved, showing that the bandwidth-bound regime and the PAop advantage are not artifacts of a single problem size.

\subsection{Ablation Study}
\label{ssec:ablation}

To quantify the per-stage contribution of the four optimizations from Section~\ref{sec:optimization}, we conduct a cumulative ablation: starting from the unmodified MFEM~v4.8 \texttt{ElasticityIntegrator} (the \emph{Baseline} row of Table~\ref{tab:ablation_study}), each row adds one further optimization on top of the previous one until PAop is reached.

All runs at $p=8$, $51.17$M~DoFs on Kunpeng with 64 MPI ranks (same partitioning as Table~\ref{tab:macro_perf_51M}); the unified GMG preconditioner gives an iteration count of $11$ for every row, so the marginal time differences reflect the cumulative effect of the operator optimizations.

\begin{table*}[htbp]
\centering
\caption{\small Cumulative ablation at $p=8$, $51.17$M~DoFs, Kunpeng 64 ranks. Stages are listed in build order (C1/C2/C3/PAop), which differs from the narrative ordering in Section~\ref{sec:optimization}.}
\label{tab:ablation_study}
\sisetup{
    table-align-text-post=false,
    output-decimal-marker={.}
}
\begin{tabular}{
    @{}
    l
    S[table-format=3.2]
    r
    S[table-format=3.2]
    r
    @{}
}
\toprule
& \multicolumn{2}{c}{\textbf{End-to-End Performance}} & \multicolumn{2}{c}{\textbf{Kernel-Level Performance}} \\
\cmidrule(lr){2-3} \cmidrule(lr){4-5}
\textbf{Optimization Stage} & {\textbf{Total Time (s)}} & {\textbf{Marg. Speedup}} & {\textbf{Kernel Time (s)}} & {\textbf{Marg. Speedup}} \\
\midrule
PA (Baseline, Section~\ref{ssec:baseline_analysis}) & 161.70 & {\textemdash} & 148.47 & {\textemdash} \\
+ Sum Factorization (C1, Section~\ref{ssec:sum_factorization_algo}) &  24.24 & $6.67\times$ &  11.65 & $12.74\times$ \\
+ Voigt Notation (C2, Section~\ref{ssec:voigt_optimization}) &  19.20 & $1.26\times$ &  11.55 & $1.01\times$ \\
+ Kernel Fusion (C3, Section~\ref{ssec:macro_kernel_fusion}) &  13.00 & $1.48\times$ &   5.38 & $2.15\times$ \\
+ Slice-wise Loops (PAop, Section~\ref{ssec:loop_optimization}) &  10.33 & $1.26\times$ & 2.74 & $1.96\times$ \\
\midrule
\multicolumn{1}{r}{Overall Speedup (vs. Baseline)} & \multicolumn{2}{c}{$15.65\times$} & \multicolumn{2}{c}{$54.19\times$} \\
\bottomrule
\end{tabular}
\end{table*}

Table~\ref{tab:ablation_study} can be read as a sequence of bottleneck shifts. Sum factorization (C1) reduces the per-element contraction cost from $O((p+1)^6)$ to $O((p+1)^4)$ and contributes the largest single step in the kernel speedup. Voigt notation (C2) alone has a small marginal effect because the FLOP and storage savings of Section~\ref{ssec:voigt_optimization} are masked by the main-memory round-trip of \texttt{QVec} from the unfused Algorithm~\ref{alg:baseline_operator_compact}. The benefit from Voigt notation becomes visible after macro-kernel fusion (C3) eliminates the round-trip and places the compact stress within the L1/L2-resident per-element working set. The final slice-wise loop reorganization (PAop) adds the cache-locality and SIMD improvements of Section~\ref{ssec:loop_optimization}.

\section{Conclusion}
\label{sec:conclusion}

High-order finite-element methods can attain high accuracy with relatively few degrees of freedom, but in practice the per-iteration cost of Full Assembly grows polynomially in $p$, and MFEM's native linear-elasticity PA path retained an $O((p+1)^6)$ element contraction, leaving the measured CPU operator-throughput sweet spot near $p \approx 2$.

This work optimized that path with sum factorization, Voigt notation, macro-kernel fusion, and slice-wise loop organization, and evaluated the resulting operator in MFEM GMG-PCG solves. The optimized operator delivers $7\text{--}83\times$ kernel and $3.6\text{--}16.8\times$ end-to-end speedup over the MFEM~v4.8 baseline across $p\in\{1,2,4,8\}$ on AMD~EPYC~7713, and the same operator-throughput sweet-spot shift to $p\ge 6$ is reproduced on Huawei~Kunpeng~920 (ARMv8.2). Ablation, FLOPs/DoF, Roofline, and cross-code bandwidth-utilization analyses explain the speedup and place PAop in context with deal.II-based elasticity codes. The optimized partial-assembly kernels, the MFEM integrator, and the GMG solver driver will be released on GitHub.

Future work will pursue three directions. First, the PAop dataflow can be transferred to other vector-valued PA operators with symmetric per-quadrature constitutive matrices. Second, the optimized tensor-product operator design can be lifted into libCEED-style backends~\cite{brown2021libceed}, enabling performance-portable physics backends across CPU and accelerator architectures. Third, the current affine tensor-product hexahedral setting can be extended to richer material models and non-affine geometries, where the balance between stored quadrature data, recomputation, and backend portability becomes more important.

\bibliographystyle{ACM-Reference-Format}
\bibliography{sample-base}

\end{document}